# Electric Field Control of Spin Orbit Coupling and Circular Photogalvanic Effect in a True Ferrielectric Crystal

Yunlin Lei[1#], Xinyu Yang[2#], Shouyu Wang[3], Daliang Zhang[4], Zitao Wang[6], Jiayou Zhang[3], Yihao Yang[1], Chuanshou Wang[1], Tianqi Xiao[3], Yinxin Bai[1], Junjiang Tian[1], Congcong Chen[8], Yu Han[7], Shuai Dong[2*], Junling Wang[5, 1*]

[1] Department of Physics & Guangdong Provincial Key Laboratory of Functional Oxide Materials and Devices, Southern University of Science and Technology, Shenzhen 518055, Guangdong, China

[2] Key Laboratory of Quantum Materials and Devices of Ministry of Education, School of Physics, Southeast University, Nanjing 211189, China

[3] College of physics and Materials Science, Tianjin Normal University, Tianjin 300387, China

[4] Multi-scale Porous Materials Center, Institute of Advanced Interdisciplinary Studies & School of Chemistry and Chemical Engineering, Chongqing University, Chongqing 400044, China

[5] Department of Physics, City University of Hong Kong, Kowloon 999077, Hong Kong SAR, China

[6] State Key Laboratory of Inorganic Synthesis and Preparative Chemistry, Jilin University, Changchun 130012, China

[7] School of Emergent Soft Matter, South China University of Technology, Guangzhou 510640, China

[8] Department of Chemistry, Southern University of Science and Technology Shenzhen, Guangdong 518055, China

*Corresponding authors: sdong@seu.edu.cn and j.wang@cityu.edu.hk

#: These authors contributed equally to this work.

Materials possessing long range ordering of magnetic spins or electric dipoles have been the focus of condensed matter research. Among them, ferri-systems with two sublattices of unequal/noncollinear spins or electric dipoles are expected to combine the properties of ferro- and antiferro-systems, but lack experimental observations in single phase materials. This is particularly true for the ferrielectric system, since the electric dipoles usually can be redefined to incorporate the two sublattices into one, making it indistinguishable from ferroelectric. This raises doubts about whether or not ferrielectricity can be considered as an independent ferroic order. Here we report the observation of true ferrielectric behaviors in a hybrid single crystal (MV)[SbBr$_5$] (MV$^{2+}$ = N,N'-dimethyl-4,4'-bipyridinium or methylviologen), where the two electric dipole sublattices switch asynchronously, thus cannot be reduced to ferroelectric by redefining the unit cell. Furthermore, the complex dipole configuration imparts circularly polarized light sensitivity to the system. An electric field can modulate the non-collinear dipole sublattices and even induce a transition from ferrielectric to ferroelectric state, thereby tuning the helicity-dependent photocurrent. This study opens a new paradigm for the study of true irreducible ferrielectricity (a new class of polar system) and provides an effective approach to the electric field control of spin-orbit coupling and circular photogalvanic effect.

**Introduction**

Long range orderings of magnetic and electric dipoles have given rise to two widely studied families of ferroic materials, and there is an almost one-to-one correspondence between them, i.e., ferromagnetism vs ferroelectricity, antiferromagnetism vs antiferroelectricity etc. However, ferrimagnetism (FiM), which exhibits an antiparallel arrangement of unequal magnetic dipoles and manifests a net magnetization, e.g., magnetite ($Fe_3O_4$) [1, 2], finds fewer counterparts in electric polar systems, except for several reports in liquid crystals [3-5]. Despite being first discussed in the 1960s [6, 7], the characteristics of ferrielectricity (FiE) remain elusive. It has been suggested that a ferri- system, FiM or FiE, should reveal the behaviors of both ferro- and antiferro- systems macroscopically, i.e., a switchable net magnetization or electric polarization and antiferro-ferro transitions. However, this has not been observed experimentally in single phase systems, even though it has been reported in composites where ferro- and antiferro- phases coexist [8-11].

In recent years, experimental and theoretical studies have uncovered some potential ferrielectric materials with non-collinearly arranged dipoles or dipoles of different magnitudes arranged in opposite directions [12-15]. Novel phenomena such as vortex domain structures, negative piezoelectricity, and chiral optical responses have been discussed [14, 16]. However, their macroscopic behavior, e.g., hysteresis loops, are exactly the same as that of ferroelectric systems (see for example, $CuInP_2S_6$ [17, 18], (MV)[$BiI_3Cl_2$] [19] and $BiCu_{0.1}Mn_{6.9}O_{12}$ [14]). This is because the polar sublattices in these materials respond to external stimuli simultaneously, thus can be reduced to one set of dipoles by redefining the unit cell. Phenomenologically, a single order parameter is sufficient to describe such systems. They are sometimes referred to as "reducible" ferrielectric materials [13]. This raises the question of whether "irreducible" FiE (or true FiE) exists or not, and what would be its distinct properties.

Initial evidence for the "irreducible" FiE in a single-phase material was discovered in $BaFe_2Se_3$ [13, 20]. In this compound, the evolution of two sets of electric dipoles upon temperature variation is asynchronous, suggesting that they cannot be reduced to one set of dipoles by doubling the unit cell and two order parameters would be needed for its phenomenological description. Unfortunately, polarization switching cannot be studied due to its high conductivity.

Here we report the observation of true "irreducible" ferrielectric behaviors in a single-phase hybrid crystal, $(MV)[SbBr_5]$. Macroscopic polarization vs electric field (P-E) measurements clearly reveal a switchable net polarization and (two) antiferroelectric-ferroelectric phase transitions. Corresponding switching current (I-E) and small-field capacitance (C-E) measurements confirm the results, which are also corroborated by first principles calculations. The complex dipole arrangement in $(MV)[SbBr_5]$ imparts chirality to it, leading to circularly polarized light sensitivity that can be tuned by external electric fields. This study opens a new paradigm in the research on hybrid polar materials with exotic dipole arrangements at atomic scale and novel functionalities.

## Results

### Structure and Optical Properties of (MV)[SbBr$_5$]

We chose the (MV)[MX$_5$] family to conduct our search for the "irreducible" FiE based on a previous study that suggests rich polar structures in this hybrid system [21]. The inorganic {MX$_5$}$_n$ chains along the *a*-axis are packed to generate the framework (Fig. 1a), wherein the MV cations are situated. In this report, we focus on (MV)[SbBr$_5$]. The as-prepared single crystals (see Methods for details) show sheet-like, plate-like, and rod-like shapes (Fig. 1b), depending on the growth conditions. Single-crystal and powder X-ray diffractions (SC-XRD & PXRD) were employed to verify phase purity and identify the (111) plane of the sheet-like single crystal (Fig. 1c).

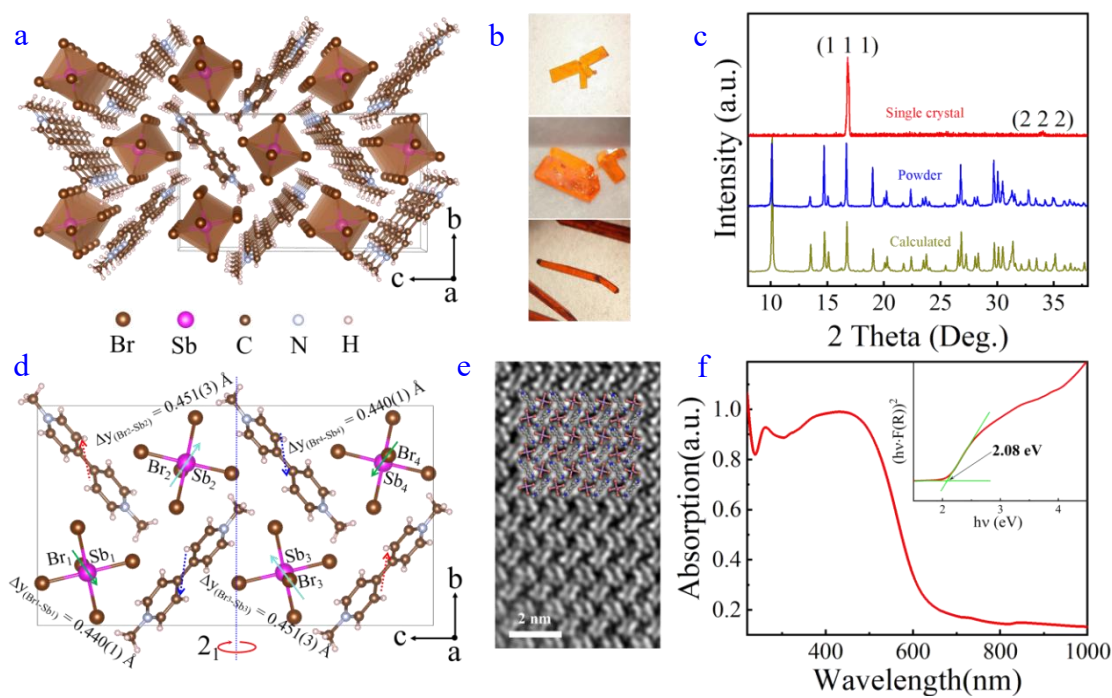

**Fig. 1 | The structure, optical photographs, XRD patterns, TEM image and absorption spectra of (MV)[SbBr$_5$].** (**a**) Schematic illustration of the structure of (MV)[SbBr$_5$] with *P*2$_1$ space group viewed along the 1D chain; (**b**) Optical

photographs of (MV)[SbBr$_5$] single crystals. (**c**) Powder, single crystal and calculated XRD patterns of (MV)[SbBr$_5$]. (**d**) The origins of electric dipoles in (MV)[SbBr$_5$]. $\Delta y_{(Br-Sb)}$ denotes the Br-Sb bond length projected along the polar *b*-axis. The arrows represent the magnitudes and directions of the dipole moments generated by the movements of Br relative to Sb and MV cations relative to the inorganic frameworks. (**e**) Low-dose high resolution TEM image of the (MV)[SbBr$_5$] structure along the [1 0 0] zone axis, superimposed with the structural model. (**f**) Absorption spectrum of (MV)[SbBr$_5$]. The inset shows the Tauc plot and the estimated direct band gap of 2.08 eV.

Detailed crystallographic analysis at room temperature reveals that the structure of (MV)[SbBr$_5$] belongs to the polar space group *P*2$_1$, with complex lattice distortions as shown in Fig. 1d. Within the inorganic framework, the Br atoms bridging the Sb atoms (Br$_{bridging}$) along the *a*-axis exhibit movements towards four different directions in the *bc*-plane, with two different shift magnitudes (indicated by the green and light blue arrows). This configuration results in two sets of electric dipoles from the inorganic chains, leading to the doubling of the unit cell and a complex polar structure. Additionally, a 2$_1$ screw axis along the *b*-axis appears, with the electric dipoles associated through the 2$_1$-axis being equal on the *bc*-plane. The clear differences in Br-Sb bond lengths (Fig. 2 and Fig. S1) illustrate the distinctions between these two sets of dipoles. To better understand the characteristics of polarization in (MV)[SbBr$_5$], we first focus on the dipoles generated by the inorganic framework. By fitting the SC-XRD data, we obtained that $\Delta y_{(Br1-Sb1)} = \Delta y_{(Br4-Sb4)} = 0.440\ (1)$ Å $\neq \Delta y_{(Br2-Sb2)} =$

$\Delta y_{(Br3-Sb3)} = 0.451$ (3) Å, where $\Delta y_{(Br-Sb)}$ refers to the projection of Br-Sb bond length within the *bc*-plane along the *b*-axis. Furthermore, we also obtained that $\Delta z_{(Sb1-Sb2)} = \Delta z_{(Sb3-Sb4)} = 5.869$ (1) Å $\neq \Delta z_{(Sb2-Sb3)} = 5.823$ (1) Å $\neq \Delta z_{(Sb1-Sb4')} = 5.915$ (1) Å, $\Delta y_{(Sb1-Sb2)} = 6.528$ (3) Å $\neq \Delta y_{(Sb3-Sb4)} = 6.541$ (6) Å $\neq \Delta y_{(Sb2-Sb3)} = \Delta y_{(Sb1-Sb4')} = 6.535$ (1) Å, where $\Delta z_{(Sb-Sb)}$ and $\Delta y_{(Sb-Sb)}$ refer to the projections of the Sb-Sb distances between neighboring [SbBr$_5$] units along the *c*-axis and *b*-axis, respectively. Note that this configuration has a synergistic impact on the MV cations, transforming it from a symmetric arrangement to an asymmetric one (Fig. S2). Furthermore, it prompts uneven displacements of MV cations along the polar *b*-axis, giving rise to additional electric dipoles. Such intricate configuration endows (MV)[SbBr$_5$] with both non-collinear and antiparallel dipoles, while exhibiting a net polarization under zero field. The later has been confirmed by second harmonic generation (SHG) measurements, as shown in Fig. S3. We have attempted to investigate the atomic displacements in (MV)[SbBr$_5$] using transmission electron microscopy (TEM). A high-resolution HAADF-STEM image along the [1 0 0] zone axis is shown in Fig. 1e and Fig. S4. However, the hybrid crystal is prone to electron beam damage, and only the intercalation of MV groups is visible but the detailed atomic displacements cannot be discerned.

Ultraviolet-visible-near-infrared (UV-Vis-NIR) absorption spectroscopy of (MV)[SbBr$_5$] at room temperature (Fig. 1f) reveals strong absorption at around 580 nm. Using the Tauc method [22], we obtained a direct bandgap of ~2.08 eV, which was further validated by theoretical calculations of the band structure (Fig. S5b).

Similar to our previous report on (MV)[SbI$_5$], this small bandgap is associated with a flat band contributed by the organic MV cations [23].

**Observation of True "Irreducible" Ferrielectricity in (MV)[SbBr$_5$]**

The multiple origins and complex arrangement of the electric dipoles revealed by the crystallographic analysis naturally lead to the question: how do the dipoles rotate/reverse under electric fields? To investigate the polarization switching characteristic of (MV)[SbBr$_5$], we deposited gold electrodes on the naturally exposed (111) planes (the crystals are too brittle to cut) and measured the P-E hysteresis loops together with the corresponding I-E curves. Immediately, we observe unique features unlike any of the known ferroelectric and/or antiferroelectric systems. As shown in Fig. 2a, at room temperature and 200 Hz, upon increasing the electric field from 0 to $E_{max}^+$, polarization increases with a corresponding broad peak in the I-E curve. However, when the electric field decreases from $E_{max}^+$ to 0, two current peaks are observed. Similar features are observed during the 0-$E_{max}^-$-0 cycle. Overall, the P-E loop shows kinks similar to that observed in antiferroelectric materials except a net polarization at zero field (see Supplementary Note S2 for further discussion), but the corresponding I-E curve is very different. Furthermore, when we decrease the temperature to -85 °C, only two peaks are observed in the I-E curve during each 0-$E_{max}$-0 cycle and the P-E loop looks just like that of an antiferroelectric system (Fig. 2b). On the other hand, when the temperature is increased to 60 °C and the frequency reduced to 20 Hz, the broad peak during the 0-$E_{max}$ scan splits into three, resulting in a total of five peaks in the I-E curve and kinks clearly visible in the P-E loop during each 0-$E_{max}$-0 cycle (Fig. 2c). Similar behaviors are also observed when we change the measurement frequency, with less switching peaks at high frequency and more at low frequency (Fig. 2d and Fig. S6). We also observed the same polarization reversal characteristics by applying electric field along the polar axis (*b*-axis) with asymmetrically arranged electrodes on the (111) surfaces (Fig. S7).

What causes such a complex polarization switching characteristics? In general, hybrid ferroelectrics exhibit slower polarization switching dynamics due to the relative bulky organic groups, which require sufficient time to rotate, translate, and/or twist. In contrast, the polarization reversal in inorganic ferroelectrics can occur in a much shorter time. Hence, we speculate that the unique switching characteristics are likely associated with the dipole reversal contributed separately by the movements of the MV groups and the Br ions. Furthermore, since the switching at ~35 kV/cm is only observed at low frequencies, it likely corresponds to the polarization reversal of the MV groups.

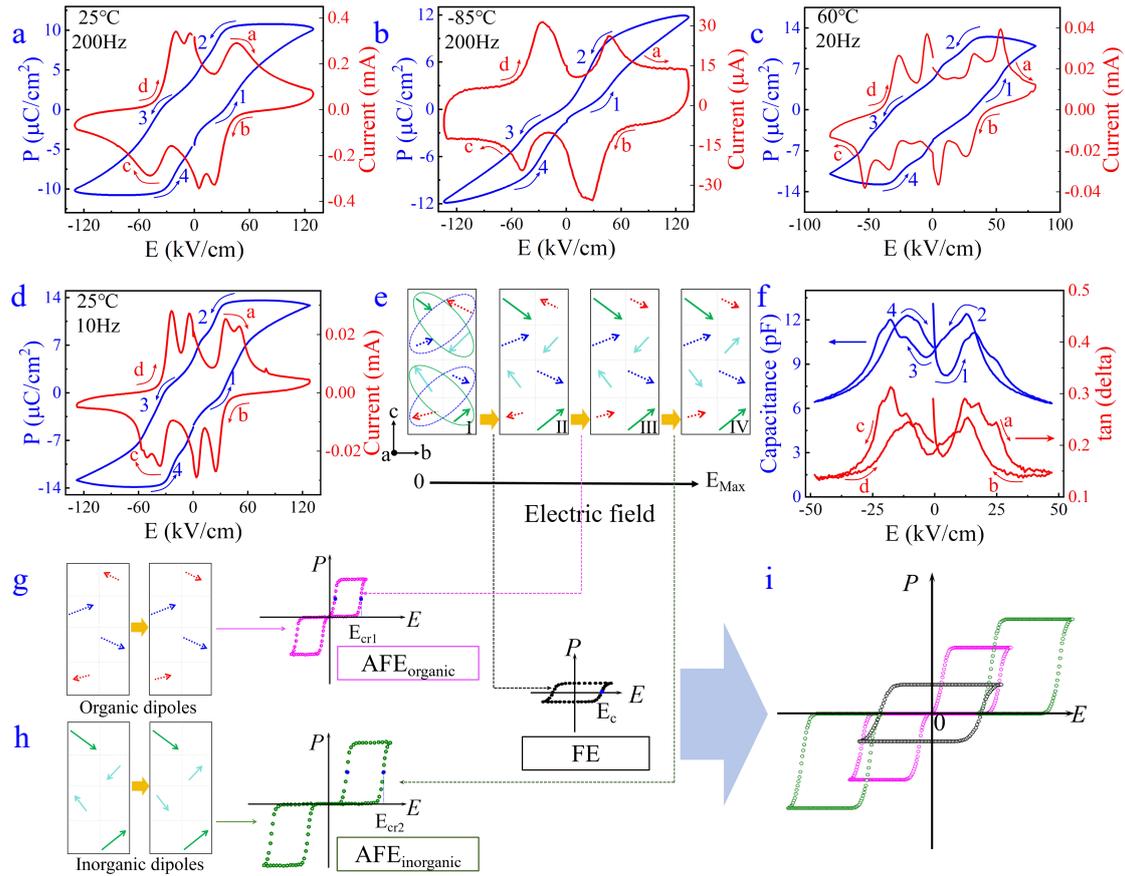

**Fig. 2 | Observation of true "irreducible" ferrielectric behaviors in (MV)[SbBr$_5$].** P-E loops and corresponding I-E curves at (a) room temperature and (b) -85 °C measured at 200 Hz, (c) P-E loops and corresponding I-E curves at 60 °C measured at 20Hz; (d) P-E loops and corresponding I-E curves at room temperature measured at 10Hz; (e) The proposed polarization reversal pathway under electric field: FiE$^{(-)}$ → FiE$^{(+)}$ → FE1 → FE2 (I → II → III → IV), the dashed and solid arrows represent the dipoles contributed by the MV groups and the inorganic frameworks, respectively. (f) Capacitance (blue) and tan δ (red) measured as functions of the DC bias superposed by a small AC driving field (~ 3.8 kV/cm) at 800 Hz at room temperature. (g) and (h), Schematics of AFE-FE transitions of the dipole sublattices from MV groups and inorganic frameworks, respectively, along with the FiE$^{(-)}$ → FiE$^{(+)}$ loops at low field.

(**i**) Superposition of the three loops leads to the P-E loop of (MV)[SbBr$_5$].

We thus propose a dipole reversal model as schematically shown in Fig. 2e, starting with the initial ferrielectric state (FiE$^{(-)}$, configuration I, the dipole pairs from the inorganic framework (solid arrows) and organic groups (dashed arrows) are marked for clarity). The corresponding configuration II with opposite net polarization is denoted FiE$^{(+)}$. Under low electric fields, no dipole flipping occurs, but the relative magnitudes of dipoles within each pair reverse, giving rise to the FiE$^{(-)}$ → FiE$^{(+)}$ hysteresis loop (black) and the small net polarizations at zero field (Fig. 2e, I-II). As electric field increases, antiferroelectric to ferroelectric (AFE-FE, Fig. 2e, II-III-IV) phase transitions occur, but with different critical fields for the organic (E$_{cr1}$, pink) and inorganic (E$_{cr2}$, green) sublattices. In fact, at a low temperature of -85 °C, this transition may not even occur for the organic sublattice, thus a single double hysteresis loop is observed in Fig. 2b and Fig. S8. As temperature increases and/or measurement frequency lowers, the ferrielectric sublattices of both the organic and inorganic origins are activated. Essentially, we observe two antiferroelectric hysteresis loops (each produces two current peaks during 0-E$_{max}$-0 scan in the I-E curve) added together, resulting in the macroscopic P-E loop and I-E curves shown in Fig. 2c and 2d, and schematically illustrated in Figs. 2g, 2h and 2i. By adjusting parameters of the FiE$^{(-)}$ → FiE$^{(+)}$ loop and the two antiferroelectric loops, we can qualitatively reproduce the P-E loops observed under different conditions (Fig. S9).

The complex electric dipole switching process is also reflected in the small signal C-E response of the single crystal (Fig. 2f). At an AC frequency of 800 Hz with DC bias sweeping from -48 kV/cm to 48 kV/cm at room temperature. The C-E curve shows three peaks in this process, similar to the I-E curve measured at high temperature and low frequency (Fig. 2c). This is because the small signal C-E measurement is highly sensitive to the instabilities of the dipole sublattices, and the DC sweeping is conducted very slowly. Under relatively small electric fields, the initial FiE state becomes unstable, and subsequent polarization reversal of the net polarization occurs. With the increase of the DC electric field, the ferrielectric states contributed by MV cations and the inorganic framework destabilize sequentially, leading to AFE-FE transitions and two more peaks in the C-E curve.

**First Principles Calculations**

To corroborate the model we proposed, especially that the dipole sublattices from the inorganic and organic components switch independently, we conducted first principles calculations (Supplementary Note S4) to extract the energy barriers for the different dipole reversals. Note that when a large electric field is applied along the *b*-axis, (MV)[SbBr$_5$] evolves into a ferroelectric phase with the *P2$_1$* space group (Fig. 3b). On the other hand, when the electric field is along the *a*-axis, it evolves into a ferroelectric phase with the *Pc* space group (Fig. S10c). We thus computed the energies for four distinct dipole configurations of (MV)[SbBr$_5$], revealing that the paraelectric (Fig. 3a), *P2$_1$* and *Pc* phases possesses energies that are 720.8 meV/u.c., 5.2 meV/u.c. and 6.4 meV/u.c. higher than the ferrielectric phase (Fig. 3c), respectively. During our experiments, the electric field is applied along [111] direction due to restrictions of crystal facets. So both *P2₁* and *Pc* are possible final phases. We have attempted to determine the final phase by in-situ Raman spectra under a DC electric field, but the samples broke down under prolonged application of the DC field necessary for Raman measurements. However, since the polarization component of the *P2₁* phase along [111] obtained from DFT calculations (17.81 μC/cm², Table S2) is much closer to the experimental value (about 14 μC/cm²) as compared to that of the *Pc* phase (39.92 μC/cm², Table S2), the high-field ferroelectric phase is considered to belong to the *P2₁* space group. Why is this phase favored in our measurements requires further study, but it doesn't affect our conclusions here.

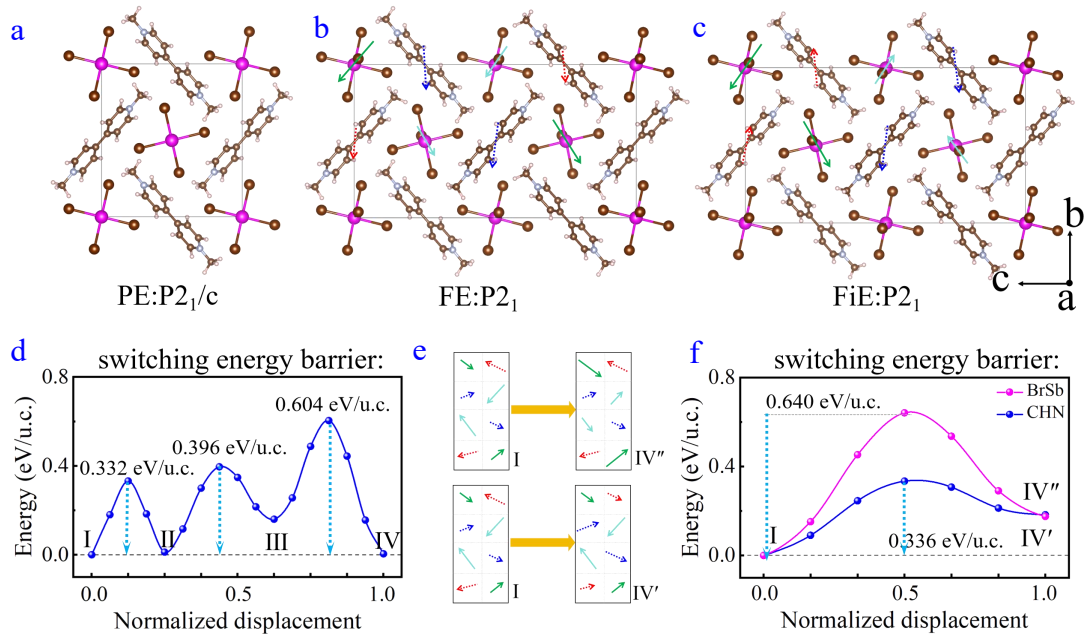

**Fig. 3 | Energy barriers for dipole reversal in (MV)[SbBr₅].** Schematic illustrations of the (**a**) nonpolar *P2₁/c* paraelectric (PE) phase, (**b**) polar *P2₁* ferroelectric (FE) phase, and (**c**) polar *P2₁* ferrielectric (FiE) phase; (**d**) Energy barriers of the dipole reversal process shown in Fig. 2e; (**e**) Schematic diagrams illustrating the dipole

reversal contributed by the inorganic frameworks and the movement of MV cations during the FiE$^{(-)}$ → FE (I → IV); (**f**) The FiE$^{(-)}$ → FE involves the polarization reversal energy barrier contributed by the inorganic framework (purple line: I → IV″) and the polarization reversal energy barrier associated with the movement of MV cations (blue line: I → IV′).

To obtain the dipole reversal barriers, we start with the initial ferrielectric state (FiE$^{(-)}$). The energy barrier for I-II transition (Fig. 3d) is 0.332 eV/u.c., much lower than that for a direct transition from FiE$^{(-)}$ to FE (0.872 eV/u.c., Fig. S11b). Notably, if we independently consider the dipole reversals associated with the inorganic framework and MV groups during the FiE$^{(-)}$→FiE$^{(+)}$→FE1→FE2 processes, we find that the barriers for MV-related dipoles are significantly lower than that for inorganic framework-related dipoles. For example, a direct FiE$^{(-)}$→FE1 transition, where FE1 represents a state in which only the organic sublattice has undergone the AFE-FE transition (IV′ in Fig. 3e), experiences an energy barrier of 0.336 eV/u.c.. On the other hand, a direct FiE$^{(-)}$→FE2 transition, where FE2 represents the state in which only the inorganic sublattice has undergone the AFE-FE transition (IV″ in Fig. 3e), experiences an energy barrier of 0.640 eV/u.c.. This indicates distinct critical fields for the two processes, consistent with our earlier analysis.

**Circular Photogalvanic Effect in (MV)[SbBr$_5$]**

While the first principles calculations corroborate our model of the electric dipoles switching process in (MV)[SbBr$_5$], the complex dipole structure also brings about unique functionalities. For example, the noncollinear and unequal Br shifts within the *bc*-plane not only lead to a net polarization and linear photogalvanic effect (LPGE), but also a 2$_1$-screw axis along the *b*-axis and a helical arrangement of dipoles within the *ac*-plane, showing chirality (Fig. 4a). This implies different responses to circularly polarized light and possibly electric field tunability. Furthermore, in semiconductors that lack inversion centers, spin-orbit coupling (SOC) splits the spin-degenerate bands, resulting in the Rashba-Dresselhaus effect [24, 25]. The two Rashba bands have opposite spins, so light excites spin up carriers preferably for a given helicity, spin down for the opposite helicity [26-28], giving rise to circular photogalvanic effect (CPGE) [29, 30]. (MV)[SbBr$_5$], with the presence of heavy element Sb and the chiral dipole configuration, offers an ideal platform for studying circularly polarized light response. On one hand, low electric fields may change the angle between the non-collinear dipoles, thus modulating the SOC. On the other hand, large electric fields induce FiE-FE transitions, which is expected to significantly increases the net polarization and the SOC. Both can be revealed by in-situ CPGE measurements.

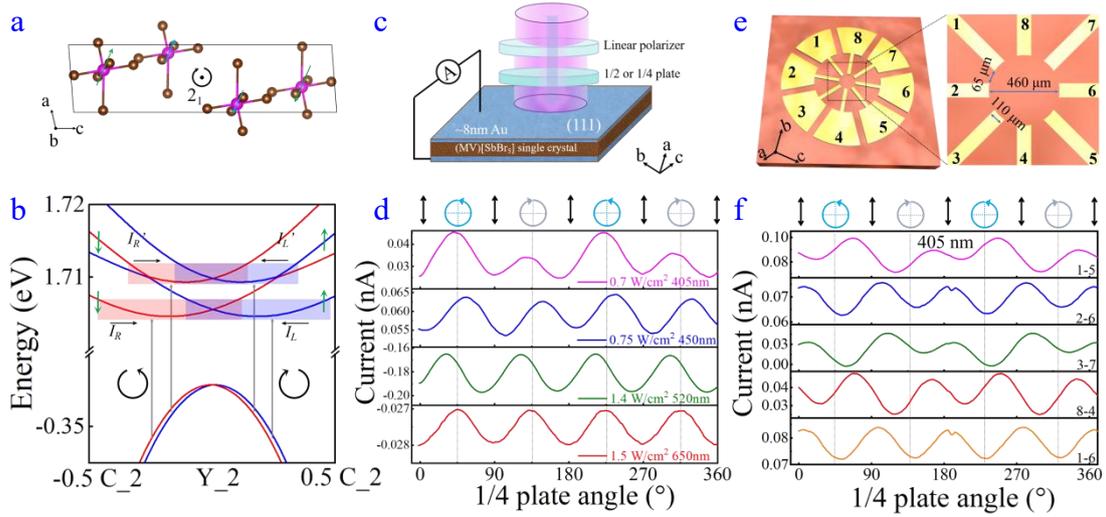

**Fig. 4 | Circular photogalvanic effect in (MV)[SbBr$_5$].** (**a**) Top view of the inorganic sublattice showing the 2$_1$-screw axes perpendicular to the inorganic chains. (**b**) Spin-polarized band structure of (MV)[SbBr$_5$] along the C_2-Y_2 direction with SOC considered; (**c**) The experimental setup to measure the linear and circular photogalvanic effects. (**d**) The room temperature photocurrent in (MV)[SbBr$_5$] crystal under lights of different wavelengths vs. λ/4 plate rotation angle; (**e**) The schematic diagram of the device for the measurement of the anisotropic CPGE effect. (**f**) Photocurrent vs. λ/4 plate rotation angle for electrodes with different numbers.

First principles calculations indeed confirm the spin polarized band structure schematically shown in Fig. 4b (Fig. S14). The spin splitting of the valence band primarily occurs perpendicular to the polarization direction, while the spin splitting of the conduction band mainly occurs along the polarization direction. It is important to note that in (MV)[SbBr$_5$], the conduction band minimum is mainly determined by the organic cation MV (Fig. S5), with a minor contribution from the inorganic framework, resulting in the absence of significant spin splitting (Fig. S14c). To observe CPGE, higher-energy photon excitation would be required.

Plate-shape single-crystals of (MV)[SbBr$_5$] (about 100 μm thick) similar to that used in the ferrielectric characterizations were coated with approximately 8 nm of gold electrodes on the (111) surfaces for photoelectric measurements. Light was vertically incident on the (111) surface, and half-wave and quarter-wave plates were used to modulate the light polarization. Cosine oscillations of photocurrent with the polarization direction of linearly polarized light were observed under lights of four different wavelengths, demonstrating the presence of linearly polarized light sensitive bulk photovoltaic effect, i.e. LPGE under zero field (Fig. S15). A reversal of photocurrent direction (for the same electric polarization direction) when the light wavelength goes below 520 nm indicates that under 520 nm and 650 nm light excitations, electrons are excited to the conduction band contributed by MV cations, whereas under 405 nm and 450 nm light excitations, electrons are excited to the higher band contributed by the inorganic framework. This is consistent with that observed in (MV)[SbI$_5$] [23]. As shown in Fig. 4d, when a quarter-wave plate is used, there is no difference in photocurrent when going from right-handed (RCP) to left-handed (LCP) circular polarization under 520 nm and 650 nm lights, while under 405 nm and 450 nm light excitations, a significant CPGE signal was detected (differences between photocurrents under LCP and RCP lights). This is consistent with the fact that, the conduction band contributed by MV cations does not exhibit spin splitting (Fig. S14c), resulting in photocurrent independent of the helicity of circularly polarized light. On the other hand, the higher band contributed by the inorganic framework exhibits significant spin splitting, thus giving rise to uneven numbers of electrons with opposite momenta when excited by RCP or LCP light.

A set of in-plane electrodes as shown schematically in Fig. 4e were prepared to investigate the correlation between photocurrent and the spontaneous polarization of (MV)[SbBr$_5$]. As shown in Fig. 4f, the photocurrent helicity sensitivity, i.e. the photocurrent difference between RCP and LCP illuminations, reverses between measurements along 1-5 and 3-7 electrode pairs. This is consistent with the fact that spin splitting occurs in the valence band along the *a*-axis and conduction band along the *b*-axis, respectively, which leads to opposite helicity-dependent photocurrents (Fig. S16). When measured along 1-6 electrode pair, the photocurrent of under RCP and LCP become nearly identical because the effects of spin splitting along *a*- and *b*-axis nearly cancel out and there is no spin splitting along the c-axis.

The phenomenological description of light polarization-dependent photocurrent can be as following [31, 27]:

$$I_{pc} = C \sin 2\varphi + L \sin(4\varphi + \varphi_0) + D \cos(4\varphi + \varphi_0) + A \quad (1)$$

where *C*, *L*, *D*, and *A* represent the amplitudes of circular photocurrent, the linear bulk photovoltaic current, the linear photon drag effect current caused by momentum transfer from photons to electrons [32], and the offset caused by other effects such as photothermal effects etc., respectively. $\varphi_0$ is to compensate for some misalignments in optics. Under zero bias (Fig. S17), the values of *C*, *L*, *D*, and *A* are 9.52 pA, 10.6 pA, 0.8 pA, and 298 pA, respectively.

When subjected to low external electric fields, the photocurrents reveal clear changes, as shown in Fig. 5a. Under 405nm laser irradiation, a change in the direction of current occurs below -0.4 kV/cm of external bias (close to the open circuit voltage: -4V, Fig. S18b), and the shape of the current-light polarization dependence curve also changes. Above -0.4 kV/cm, the RCP photocurrent is greater than the LCP photocurrent, while below -0.4 kV/cm, the RCP photocurrent is smaller than the LCP photocurrent. Fig. 4b shows the variation of the four parameters *C*, *L*, *D*, and *A* fitted according to Eq. (1) under different external electric fields (See Supplementary Note S6 for detailed discussion). It should be noted that under different fields, the direction of the current due to circular BPGE remains unchanged. Under short-circuit conditions ($E = 0$), the red and blue square regions in the Fig. 4b can be occupied by electrons excited by RCP and LCP lights, respectively. Under bias, external bias can change the distribution of available state density in *k*-space, thereby changing the magnitudes of $I_{ph}^{R}$ and $I_{ph}^{L}$ [26, 28].

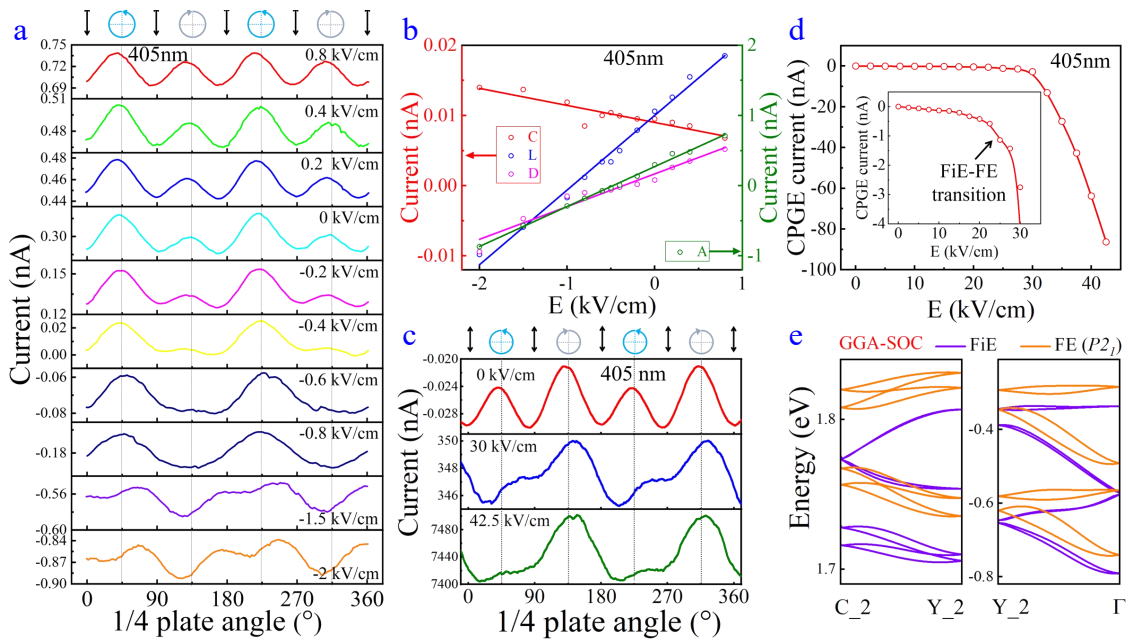

**Fig. 5 | Circular photogalvanic effect in (MV)[SbBr$_5$] under low and high electric

**fields.** (**a**) Photocurrent versus the λ/4 plate angle under different external bias. (**b**) The different contributions to the photocurrent. (**c**) Photocurrent versus the λ/4 plate angle under low and high bias. (**d**) The external electric field dependence of CPGE photocurrent, inset: a magnified view between 0 and 30 kV/cm. (**e**) Comparison of the bands between the ground state FiE and the electric-field-induced ferroelectric state ($P2_1$ phase) considering SOC.

The change in CPGE under low fields can be understood if we consider that the dipole sublattices and the net polarization are changed by the external field, which then affect the spin splitting of the bands, as shown in Fig. S20a. An increasing polarization causes the points $-k_1$ and $k_2$ to shift along $-k_y$, resulting in an increase in the net photocurrent $I_L$. Consequently, the CPGE photocurrent coefficient $C = 2|I_L|$ also increases. Conversely, when a bias opposite to the polarization direction is applied, the polarization decreases, and C decreases as well. With ΔC representing the compensation of the CPGE photocurrent coefficient due to the change in polarization under bias, the CPGE photocurrent coefficient $C_{exp}$ should decrease with increasing field, which is consistent with the experimental observations. A detailed discussion on the microscopic mechanism of linear photon drag effect and the changes in the photocurrent under an external bias is presented in the Supplementary Note S7 [33, 34].

The above analysis implies that, with higher fields inducing a transition from the ground-state FiE to the ferroelectric state in (MV)[SbBr$_5$], a significant increase in polarization will further strengthen the SOC, and a substantial impact on the CPGE current should be expected. We thus prepared thinner (approximately 40 μm thick) plate-shaped (MV)[SbBr$_5$] single crystals and repeated the measurements. As shown in Fig. S23, the current-electric field curve measured reproduces the polarization reversal characteristics of the thicker samples. By applying an electric field of 0 kV/cm→ 50 kV/cm → 0 kV/cm, the thin (MV)[SbBr$_5$] crystal was driven into a single-domain state. Subsequently, 405 nm light was vertically irradiated onto the (111) surface of (MV)[SbBr$_5$], and a DC field was applied to measure the current under this illumination. As shown in Fig. 5c, under increasing external bias, the shape of the photocurrent underwent significant changes. Additionally, both the magnitude of the photocurrent and the difference between LCP and RCP light illuminations are dramatically altered. The helicity-dependent photocurrent "*C*" as a function of the external electric field is shown in Fig. 5c. As the electric field increases, *C* slowly increases at first, then rises sharply around 25 kV/cm, corresponding to the current peak at 25 kV/cm in the current-electric field curve (Fig. S23), which is associated with the "FiE-FE" transition, suggesting a significantly enhanced SOC in the electric field-induced FE phase.

    As show in the Fig. 5e and Fig. S24, first principle calculations confirm that the spin-splitting in the valence bands along the "Γ-A" and "Y_2-Γ" paths, as well as in the conduction bands along the "A-E" and "C_2-Y_2" paths, are significantly enhanced when the system transitions from the FiE state to the *P2$_1$* FE state. When the excited electrons are primarily governed by spin splitting in the conduction band, as

shown in Fig. S25a, under LCP light, the current generated by photoelectrons excited at $-k_1$ increases during the relaxation process, while that excited at $k_2$ decreases, leading to an overall increase in the net photocurrent. The same is true when the excited electrons are governed by valence band as well (Fig. S25b).

**Conclusions**

In summary, we report the observation of true "irreducible" ferrielectric behaviors in a hybrid crystal (MV)[SbBr$_5$] with complex non-collinear dipole arrangements. Temperature and frequency-dependent measurements indicate that the electric dipoles from the MV groups and the inorganic framework in (MV)[SbBr$_5$] undergo asynchronous switchings, leading to the combination of a ferroelectric hysteresis loop and (two) AFE-FE transitions. First principles calculations confirm the model by revealing that the dipoles of MV groups have a smaller switching barrier. The complex non-collinear dipole arrangement also confers chirality to (MV)[SbBr$_5$]. We observed that the photocurrent depends on the circularly polarized light helicity and achieved modulation of helicity-dependent photocurrent through electric field. This helicity-dependent photocurrent modulation is fundamentally driven by the electric field's influence on the non-collinear dipole sublattices and the FiE-FE transition, which enhances polarization and, in turn, strengthens the SOC. This true ferrielectric order, distinct from both ferroelectric and antiferroelectric orders, offers a new paradigm for the study of ferroic systems and provides a novel approach for the electric field control of SOC and circular photogalvanic effect.

photocurrents with light polarization. *Nat. nanotechnology* **7**, 96-100 (2012).

## ACKMOWLEDGEMENTS

We acknowledge the Big Data Computing Center of Southeast University for providing computational resources.

## FUNDING

This work was supported by the Guangdong Innovative and Entrepreneurial Research Team Program (Grant No. 2021ZT09C296), National Natural Science Foundation of China (Grant Nos. 12074164, 12325401 & 12274069) and Guangdong Provincial Key Laboratory Program (2021B1212040001) from the Department of Science and Technology of Guangdong Province, China.

## AUTHOR CONTRIBIBUTIONS

J.W., S.D., and Y.L. conceived the idea and designed the study. X.Y. performed the DFT calculations. Y.L., S.W., J.Z., and T.X. conducted ferroelectricity measurements. D.Z., Z.Z. and Y.H., performed TEM measurements. Y.Y. conducted PXRD measurements. Y.L., Y.Y., and Y.B. conducted optoelectronic measurements. C.C. conducted single crystal XRD measurements. All authors contributed to the discussion of the results and writing of the manuscript.

### Conflict of interest statement

The authors declare no competing interests.

Supplementary Information for

Electric Field Control of Spin Orbit Coupling and Circular Photogalvanic Effect in a True Ferrielectric Crystal

Yunlin Lei[1#], Xinyu Yang[2#], Shouyu Wang[3], Daliang Zhang[4], Zitao Wang[6], Jiayou Zhang[3], Yihao Yang[1], Chuanshou Wang[1], Tianqi Xiao[3], Yinxin Bai[1], Junjiang Tian[1], Congcong Chen[8], Yu Han[7], Shuai Dong[2*], Junling Wang[5, 1*]

[1] Department of Physics & Guangdong Provincial Key Laboratory of Functional Oxide Materials and Devices, Southern University of Science and Technology, Shenzhen 518055, Guangdong, China

[2] Key Laboratory of Quantum Materials and Devices of Ministry of Education, School of Physics, Southeast University, Nanjing 211189, China

[3] College of physics and Materials Science, Tianjin Normal University, Tianjin 300387, China

[4] Multi-scale Porous Materials Center, Institute of Advanced Interdisciplinary Studies & School of Chemistry and Chemical Engineering, Chongqing University, Chongqing 400044, China

[5] Department of Physics, City University of Hong Kong, Kowloon 999077, Hong Kong SAR, China

[6] State Key Laboratory of Inorganic Synthesis and Preparative Chemistry, Jilin University, Changchun 130012, China

[7] School of Emergent Soft Matter, South China University of Technology, Guangzhou 510640, China

[8] Department of Chemistry, Southern University of Science and Technology Shenzhen, Guangdong 518055, China

*Corresponding authors: sdong@seu.edu.cn and j.wang@cityu.edu.hk

#: These authors contributed equally to this work.

**Note S1: Materials and Methods**

**Materials:** Antimony tribromide (SbBr$_3$, Alfa Aesar), 4,4′-bipyridine (Mreda), Hydrobromic acid (HBr, 48%, Macklin), Methanol (MeOH, Xilong scientific). All the chemicals were bought and used without further purification.

**Synthesis of (MV)[SbBr$_5$]:** The (MV)[SbBr$_5$] were synthesized by using solvothermal reaction method with bipyridine radical cations molecular and metal halide compound. Firstly, 0.12 g of SbBr$_3$, 0.8 ~ 0.9 mL of HBr (48%), 0.052 g of 4,4′-bipyridine, and 10 mL MeOH were mixed in a 25 mL Teflon bomb. The Teflon bomb was then sealed in a Parr autoclave and heated in a programmable oven with the following parameters: heating from 25 °C to the designated temperature of 150 °C, holding at 150 °C for 30 h, and then cooling to 25 °C with 2 ~ 4 °C/h. This process yielded mostly rod-shaped crystals with a smaller amount of sheet- and plate-shaped crystals. A larger quantity of HBr and a faster cooling rate led to more sheet-shaped crystals.

**Structural characterizations:** Variable-temperature single-crystal diffraction data of (MV)[SbBr$_5$] were collected on a Rigaku XtalAB PRO MM007DW diffractometer with Mo-K$\alpha$ radiation ($\lambda$ = 0.77 Å) at 200 K and room-temperature, respectively. Data were collected and reduced using the Bruker APEX3 program. The structures were determined by a direct method and refined with the OLEX2 program package based on $F^2$ with refinements of full-matrix least squares. For the structure at room temperature, $R_1$:3.2%, $wR_2$:7.81%, Goof:1.09. Powder X-ray diffraction (PXRD) for (MV)[SbBr$_5$] was performed on a Rigaku D-max 2500 PC with Cu-Kα radiation at room temperature. Ultralow-dose high-resolution transmission electron microscopy (HRTEM) was conducted using a double-Cs-corrected Spectra 300 transmission electron microscope, operated at an acceleration voltage of 300 kV. The HRTEM images were acquired with a Gatan K3 direct-detection camera, utilizing electron-counting mode to maximize detection efficiency. To minimize the structural damage, the electron dose was meticulously regulated to remain < 20 e/Å$^2$. Furthermore, an averaged background filter was employed to augment the signal-to-noise ratio, thereby enhancing image quality. Ultraviolet-visible-near-infrared (UV-vis-NIR) absorption spectra measurement was performed at room temperature using a UV-3600 UV-vis-NIR spectrophotometer.

**Electrical characterizations:** For ferroelectric measurements, gold electrodes of approximately 15 nm thick were deposited onto the (111) planes of plate-shaped crystals using magnetron sputtering. The polarization-electric field (P-E) hysteresis loops, current-electric field curves and capacitance-electric field (C-E) response were measured by TF Analyzer 3000. For temperature dependent measurements, the samples were cooled using liquid nitrogen, with temperature control facilitated by the TCU 3016. Photoelectric measurements were performed with a parallel electrode configuration. The light source consisted of laser diodes (KYD405NX-T1685, KYD450N100-T1685, KYD520N100-G2290 and D650NX-T1685, Shenzhen Jukun Optical Technology Co., Ltd, China.) emitting light at wavelengths of 405 nm, 450 nm, 520 nm and 650 nm, respectively. The photocurrent response under different laser illuminations were measured using Keithley 2636B. For the gold electrodes grown on the (111) plane to measure the photocurrent in Fig. 4d, Fig. 5a, and Fig. 5c, the areas of the gold electrodes are 1 mm², 6.2 mm², and 1.8 mm², respectively.

**Density functional theory (DFT) calculations:** The density functional theory (DFT) calculations were carried out using projector-augmented wave pseudopotentials as implemented in the Vienna *ab initio* Simulation Package (VASP) [1]. The exchange-correlation functional was treated using Perdew-Burke-Ernzerhof (PBE) parametrization of the generalized gradient approximation (GGA) [2]. More tests with different exchange-correlation functionals (PBEsol) [3] can be found in Table S1, in comparison with the experimental ones. The van der Waals (vdW) correction of the DFT-D3 method was applied [4]. The energy cutoff was fixed to 500 eV and the k-point grids of 6×3×2 were adopted for both optimization and static calculation. The convergent criterion for the energy was set to $10^{-6}$ eV, and the criterion of the Hellman-Feynman force during the structural relaxation was 0.01 eV/Å for all atoms. To obtain more accurate band gaps, the hybrid Heyd-Scuseria-Ernzerhof (HSE06) functional was employed, and the energy convergent criterion was set to $10^{-4}$ eV for the HSE06 functional [5]. The most likely switching pathways among different transition states were evaluated using the nudged elastic band (NEB) method [6]. The theoretical value of ferrielectric polarization was estimated to be 0.29 μC/cm² along the *b*-axis based on the standard Berry phase method [7]. Additionally, we have estimated the local dipole moments of every

(MV)[SbBr$_5$] pairs (A, B, C and D) along the *b*-axis, as illustrated in Fig. S10. Every pairs contain one [MV]$^{2+}$ molecule and one [SbBr$_5$]$^{2-}$ group. As summarized in Table S2, the dipole along *b*-axis of pair A (D) is 12.72 |e|Å, while the neighboring pair B (C) contributes -12.53 |e|Å. The net dipole of one unit cell after compensation is 0.38 |e|Å along the *b*-axis (corresponding to $P_b$=0.33 µC/cm$^2$, very close to aforementioned net polarization).

**Note S2: Further analysis of the remnant polarization.**

There is a substantial remnant polarization after the application of an electric field under ambient condition, as shown in Fig. 2a and 2d. This is attributed to the close energy proximity between the ferroelectric and ferrielectric phases, as confirmed in the DFT section. By increasing the measurement frequency, reducing the driving electric field, or lowering the temperature, the remnant polarization at zero field decreases (Fig. 2b and S6), approaching the calculated net polarization value (~0.33 µC/cm$^2$ along *b*-axis ) of the ground ferrielectric phase. This indicates that after the application of the electric field results in a transition from ferrielectric to ferroelectric phases, the structure can revert to the ferrielectric phase once the electric field is removed, though this recovery may require some time due to the proximity of their energies.

**Note S3: Characterization of polarization reversal features under an electric field applied along the b-axis.**

We applied asymmetric electrodes using silver paste on the two (111) surfaces of the plate-like (MV)[SbBr$_5$] crystals, with the connection between the two electrodes approximately along the b-axis, as shown in Fig. S7a. At 35°C, we observed P-E loops with features similar to those observed when the electric field was applied along the [111] direction. We also observed the same frequency dependence as shown in Fig. S7 b-e. After performing 30 consecutive P-E loop measurements at the same frequency, the shape of the P-E loops remained unchanged, indicating the recovery of the ground FiE state after the AFE-FE transition.

**Note S4: DFT and further analysis of the polarization reversal pathways.**

By rotating the dipoles in the ground-state FiE phase and optimizing the structure using DFT, we obtained two possible ferroelectric phases with *P2$_1$* and *Pc* space groups corresponding to when the electric field is applied along *b*-axis and *a*-axis respectively, as shown in Fig. S10b and S10c. To evaluate the polarization of the local dipoles, we define one MV$^{2+}$ molecule in conjunction with one [SbBr$_5$]$^{2-}$ group as forming an uncharged pair. The local dipole moments of each (MV)[SbBr$_5$] pair (A, B, C, and D) are estimated along *a*, *b* and *c*-axis, as illustrated in Fig. S10. As summarized in Table S2, the total dipoles per unit cell are calculated to be (0, 0.33, 0), (0, -43.73, 0), and (-18.63, 0, -81.35) μC/cm² for the FiE phase, the *P2$_1$* FE phase, and the *Pc* FE phase , respectively. These values are close to values obtained using the standard Berry phase method, which were computed as (0, 0.29, 0), (0, -43.96, 0), and (-18.48, 0, -81.64) μC/cm². The polarization values of four (MV)[SbBr$_5$] pairs of the intermediate phase, corresponding to the AFE-FE transition occurring only in the organic part, have also been calculated and listed in Table S2. However, since the AFE-FE transition in the organic part is expected to influence the inorganic part, these values may differ slightly from the actual ones. Furthermore, since the experimentally observed polarization value when the electric field is applied along [111] direction is close to that of the $P2_1$ FE phase when projected onto [111] direction (Table S2), we conclude that $P2_1$ FE phase is the high field FE phase in our study.

As shown in Fig. S12, when the electric field is applied along the *a*-axis, the dipoles that are initially aligned opposite to the electric field gradually orient in the direction of the field. First, the AFE-FE transition of the organic dipoles occurs, followed by the AFE-FE transition of the inorganic dipoles. Meanwhile, the net polarization disappears along the *b*-axis and appears along both the *a*-axis and *c*-axis.

**Note S5: Helicity-sensitive photocurrent.**

As shown in Fig. S18a, under the 405nm laser irradiation, the photocurrent under RCP light is significantly higher than that under LCP light. The anisotropy factor ($g_{Iph}$) is calculated by the following equation: $g_{Iph} = 2(I_{ph}^R - I_{ph}^L)/(I_{ph}^R + I_{ph}^L)$, $I_{ph}^R$ and $I_{ph}^R$ are the photocurrents under RCP and LCP lights, respectively. The measured value of $g_{Iph}$ under 405 nm laser irradiation is 0.064.

**Note S6: The microscopic mechanism behind the change in the CPGE photocurrent.**

We seek to understand the microscopic mechanism of CPGE to explain the changes of the photocurrent under external bias. For CPGE, the process is enabled by converting the angular momentum of photons into the translational motion of charge carriers [8]. As shown in Supplementary Fig. 19a, under excitation by RCP/LCP light at a frequency of ω, energy and momentum conservation permit electron transitions at two different $k_y$ values. During the relaxation of photo-excited electrons, the asymmetry momentum distribution of electrons results in a net short-circuit currents $I_R$ or $I_L$. RCP/LCP light have opposite angular momenta, thus the CPGE currents generated by RCP/LCP light have the same magnitude but opposite directions. As shown in Supplementary Fig. 19b and Supplementary Fig. 19c, assuming the bias does not alter the intrinsic band structure of the material and neglecting spin splitting in the valence band, and $C_0$ denote the CPGE photocurrent coefficient under an external bias. Under an external bias, the excited electrons gain additional momentum, altering their distribution in the conduction band. This causes the photocurrents $I_R$ and $I_L$, generated under RCP/LCP light excitation, to increase or decrease by equal amounts. Since the CPGE coefficient $C_0$ is defined as half the difference between the photocurrents under right and left circularly polarized light, i.e., $C_0 = (I_R - I_L)/2$, we expect $C_0$ to not change sign, consistent with our experimental observations.

**Note S7: The microscopic mechanism behind the change in the LPDE photocurrent under an external electric field.**

Regarding the effect of bias on LPDE, LPDE is generated by the linear momentum transfer from photons to charge carriers [8, 9]. For simplicity, we ignore SOC-bands in the LPDE analysis (Fig. S21 a-d). As shown in Fig. S21a, photons carrying in-plane momentum (indicated by tilted dashed arrows) can create non-uniform momentum distribution (from $k_3$ to $k_4$) of photoexcited electrons in the conduction band, with net momentum marked by red rectangles. Under bias, the distribution of excited electrons changes, which affects the net momentum. At higher negative bias, this results in a sign change of the LPDE photocurrent, aligning with our experimental observations. Although we have provided a physical explanation for CPGE and LPDE under bias, the linear photogalvanic effect under bias requires and awaits a more comprehensive theoretical treatment.

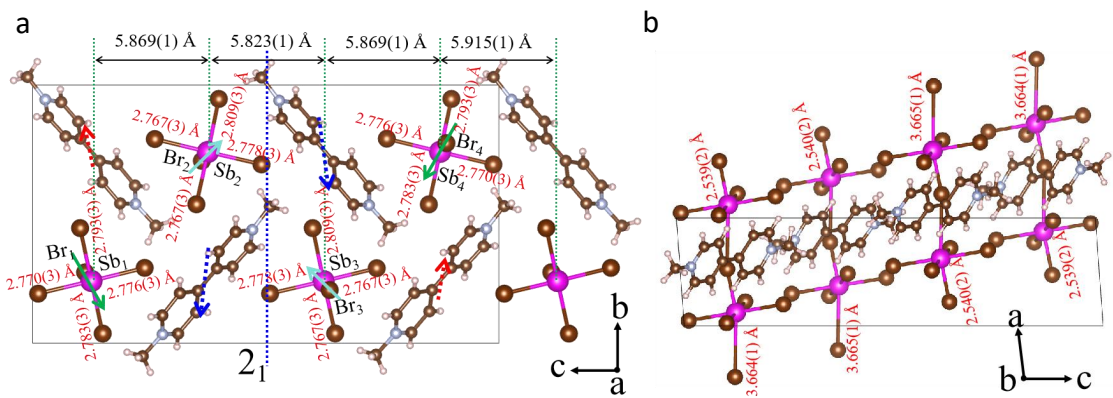

Fig. S1. Bond lengths within the inorganic framework of (MV)[SbBr$_5$]. a, The distances between the inorganic frameworks and the organic groups along the *c*-axis and the lengths of the lateral Br-Sb bonds. b, The lengths of the axial Br-Sb bonds in the inorganic framework.

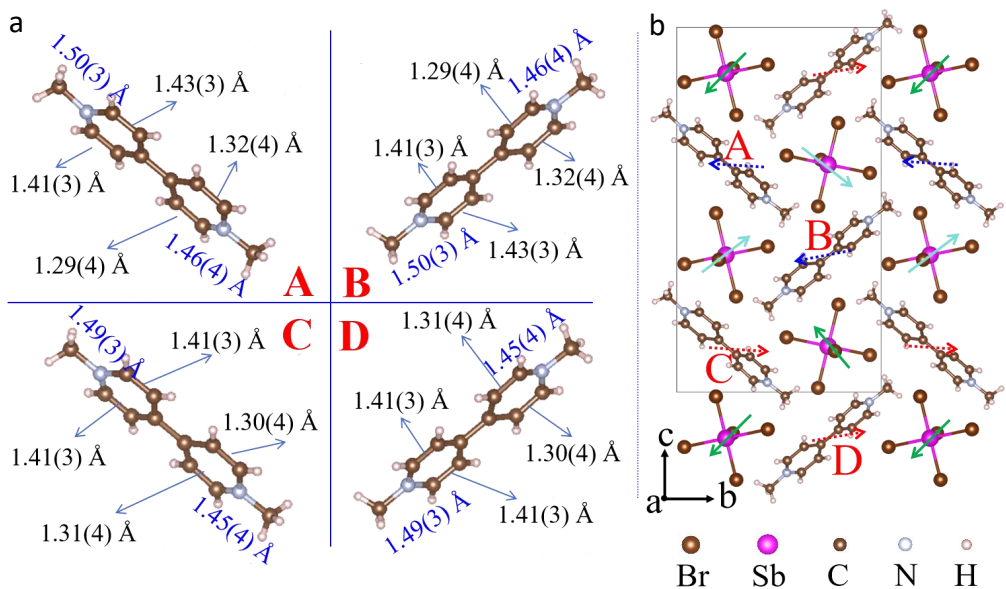

Fig. S2. Distortions of the MV groups in (MV)[SbBr$_5$]. a, The various bond lengths in the methylviologen (MV) cations, indicating the transition from a symmetric structure to an asymmetric structure. b, The locations of the MV groups in the crystal.

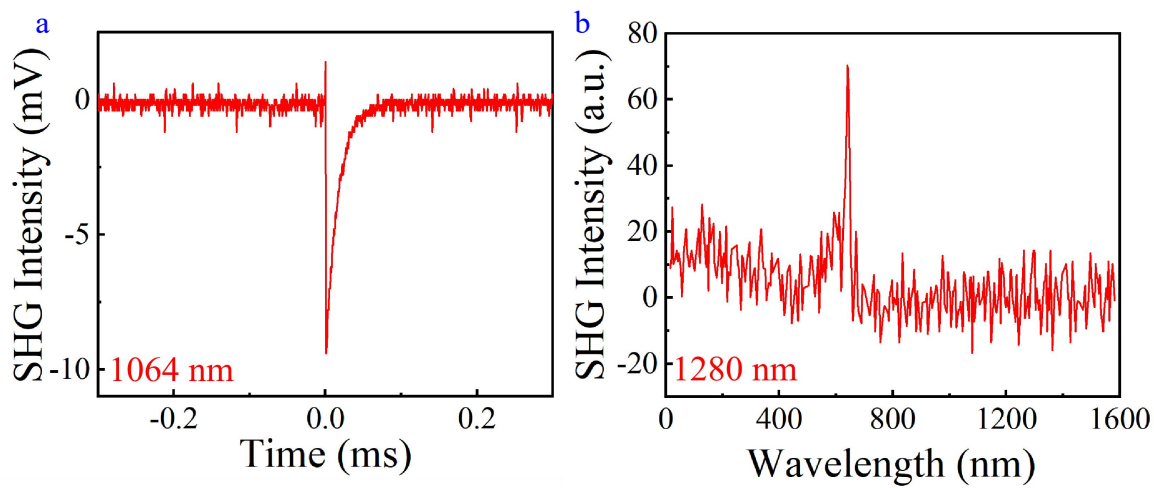

Fig. S3. The SHG of the (MV)[SbBr$_5$] crystal. a, The SHG signals of (MV)[SbBr$_5$] were measured by the Kurtz-Perry method on a Q-switched Nd:YAG solid-state laser with 1064 nm at room temperature, crystals were ground and sieved into a particlesize range 244−355 μm. b, SHG spectra of (MV)[SbBr$_5$] crystal under 1280 nm laser irradiation.

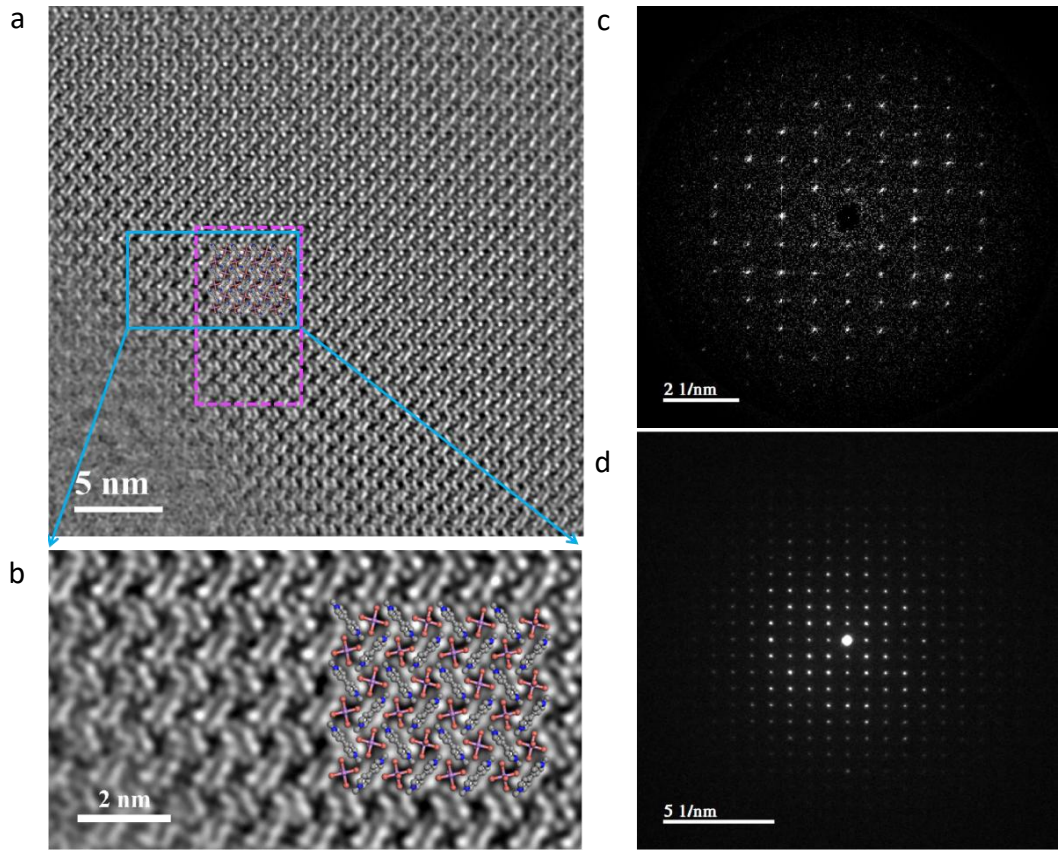

Fig. S4. TEM Characterizations of the (MV)[SbBr$_5$]. a, Low-dose high resolution TEM image of (MV)[SbBr$_5$] structure along the [1 0 0] zone axis, The magnified view of the region outlined by the pink dashed box is shown in the Fig. 1e; b, The magnified image of low-dose high resolution TEM image of the (MV)[SbBr$_5$], superimposed with the structural model. Fast Fourier Transform (FFT) of the image (c) and SAED simulations (d) of (MV)[SbBr$_5$].

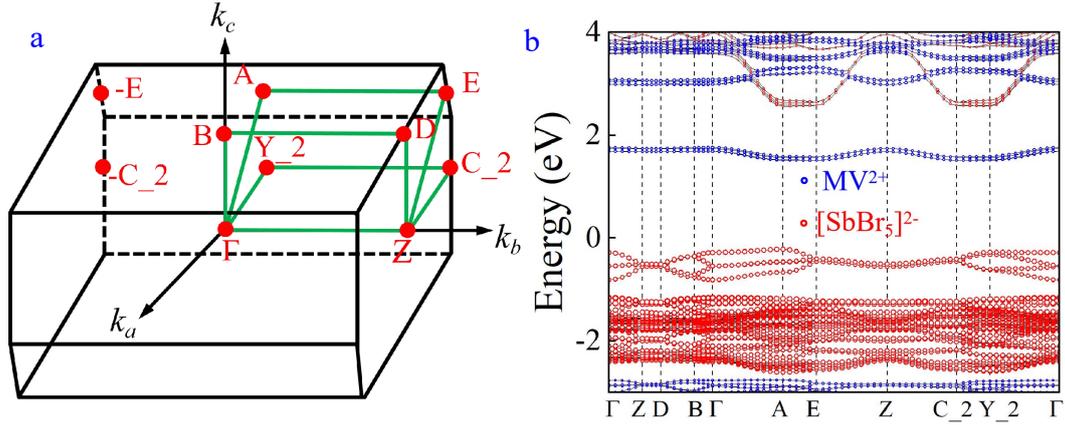

Fig. S5. Density functional theory (DFT) calculations. a, The Brillouin zone of (MV)[SbBr$_5$]; b, DFT-HSE band structure of (MV)[SbBr$_5$], which leads to a band gap ~1.73 eV.

Table S1 The structural parameters of (MV)[SbBr$_5$] at 200 K and 290 K, compared with the DFT (GGA) calculated values.

|  | Space group | a (Å) | b (Å) | c (Å) | α = γ | β | V (Å$^3$) |
|---|---|---|---|---|---|---|---|
| 200 K | $P2_1$ | 6.076 | 12.905 | 23.666 | 90° | 94.035° | 1851.106 |
| 290 K | $P2_1$ | 6.104 | 13.070 | 23.537 | 90° | 94.140° | 1872.912 |
| GGA-PBE-D3 | $P2_1$ | 6.191 | 12.768 | 23.460 | 90° | 94.691° | 1848.174 |
| GGA-PBEsol | $P2_1$ | 6.034 | 12.763 | 24.324 | 90° | 95.743° | 1863.976 |
| GGA-PBEsol-D3 | $P2_1$ | 5.969 | 12.371 | 23.425 | 90° | 94.162° | 1725.164 |

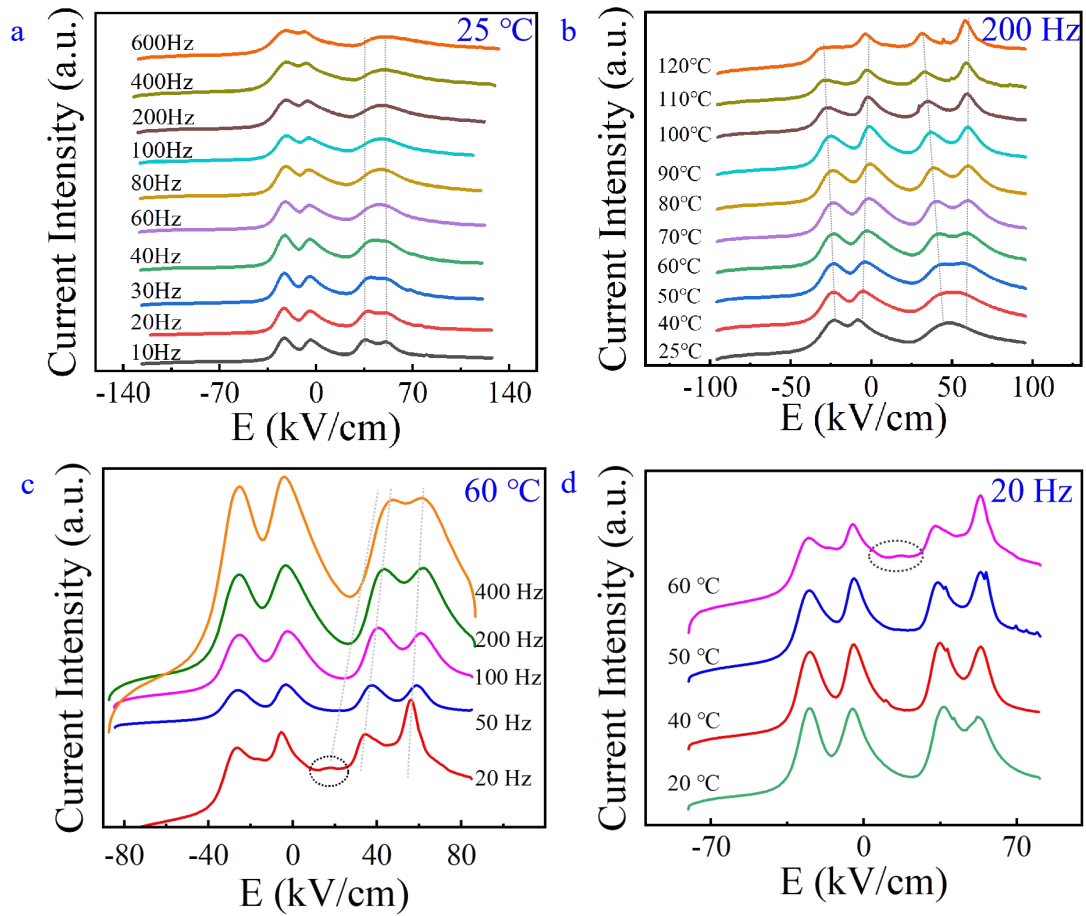

Fig. S6. Variation in the number of switching current peaks as functions of temperature and frequency. a, Variation in the number of current peaks in the first and second quadrants of current-electric field curves at different frequencies. b, Variation in current peaks at 200 Hz frequency and different temperatures, c, Variation in current peaks at 60 °C and different frequencies. d, Variation in current peaks at 20 Hz frequency and different temperatures.

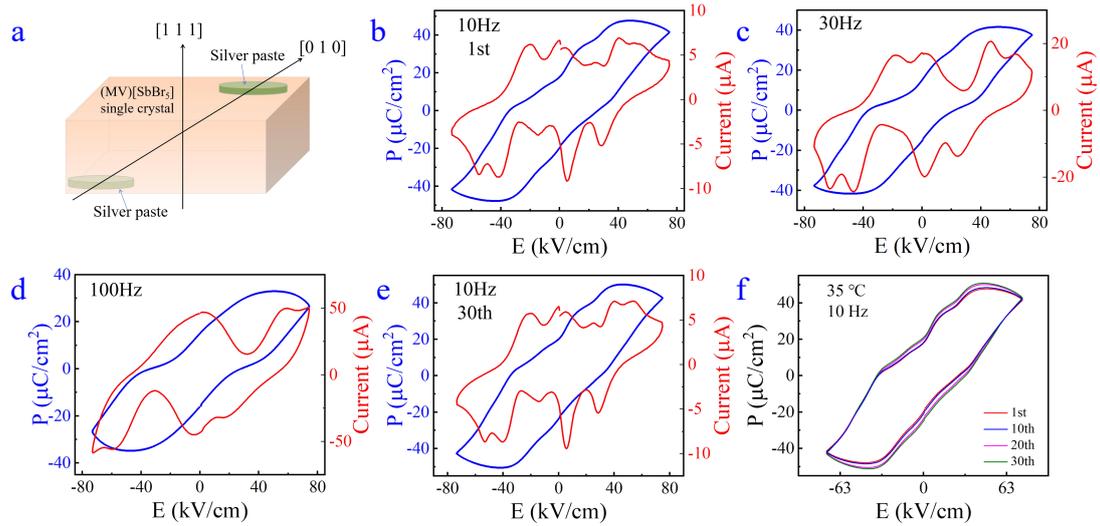

Fig. S7. Characterization of polarization reversal behavior along the b-axis (polar axis). a, Schematic diagram of polarization reversal characterization along the b-axis, where silver paste is applied to two asymmetric positions on the upper and lower (111)-plane of the (MV)[SbBr$_5$] single crystal, and the line connecting the two electrodes is approximately along the b-axis. P-E loops and corresponding I-E curves at 35 °C measured at (b) 10 Hz, (c) 30 Hz and (d) 100 Hz. e, The P-E loop and corresponding I-E curve after 30 consecutive tests at 35 °C and a frequency of 10 Hz. F, Comparison of the P-E loops during the first, 10th, 20th, and 30th tests at 35 °C and a frequency of 10 Hz.

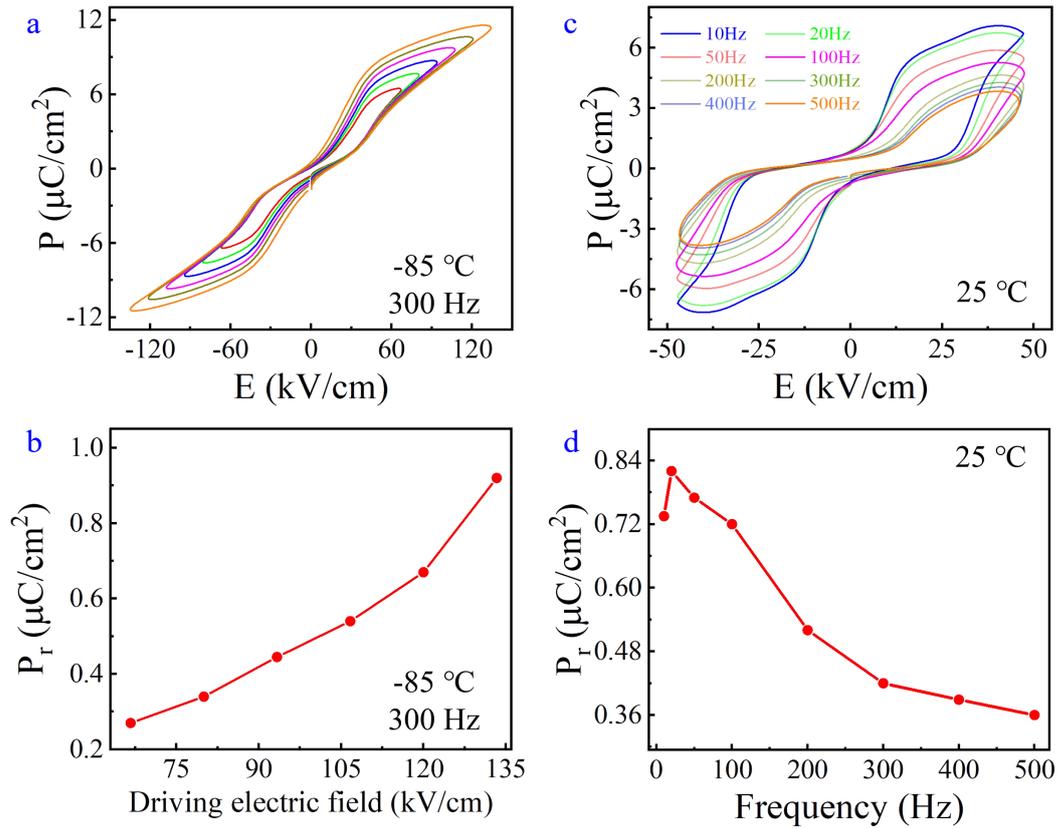

Fig. S8. Further analysis of remnant polarization. a, b, Hysteresis loops measured at -85 °C and 300 Hz under different driving electric fields and the corresponding dependence of remnant polarization along [111] on the driving electric field. c, d, Hysteresis loops measured at room temperature under low driving electric fields at different frequencies and the corresponding dependence of remnant polarization along [111] on frequency.

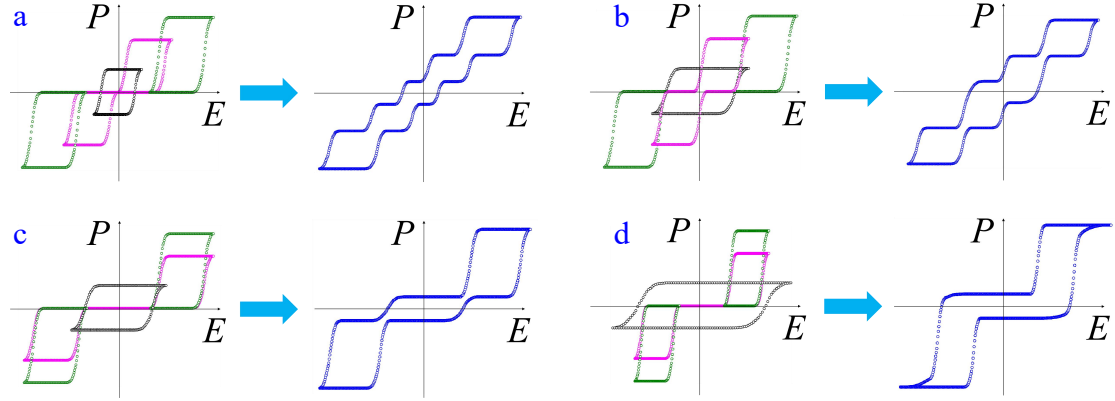

Fig. S9. Schematics of the hysteresis loop obtained with different parameters of the dipole reversal process. The numbers of current peaks during the 0 to $E_{max}$ and $E_{max}$ to 0 sweepings are denoted $m$ and $n$, respectively. The coercive field of the FE loop (black) is denoted $E_c$, and the critical fields for the AFE-FE transitions of the MV groups and the inorganic sublattices are denoted $E_{cr1}$ and $E_{cr2}$, respectively. a, $E_c < E_{cr1} < E_{cr2}$, $m = 3$, $n = 2$; b, $E_c \approx E_{cr1} < E_{cr2}$, $m = 2$, $n = 2$; c, $E_c < E_{cr1} \approx E_{cr2}$, $m = 2$, $n = 1$; d, $E_c \approx E_{cr1} \approx E_{cr2}$, $m = 1$, $n = 1$;

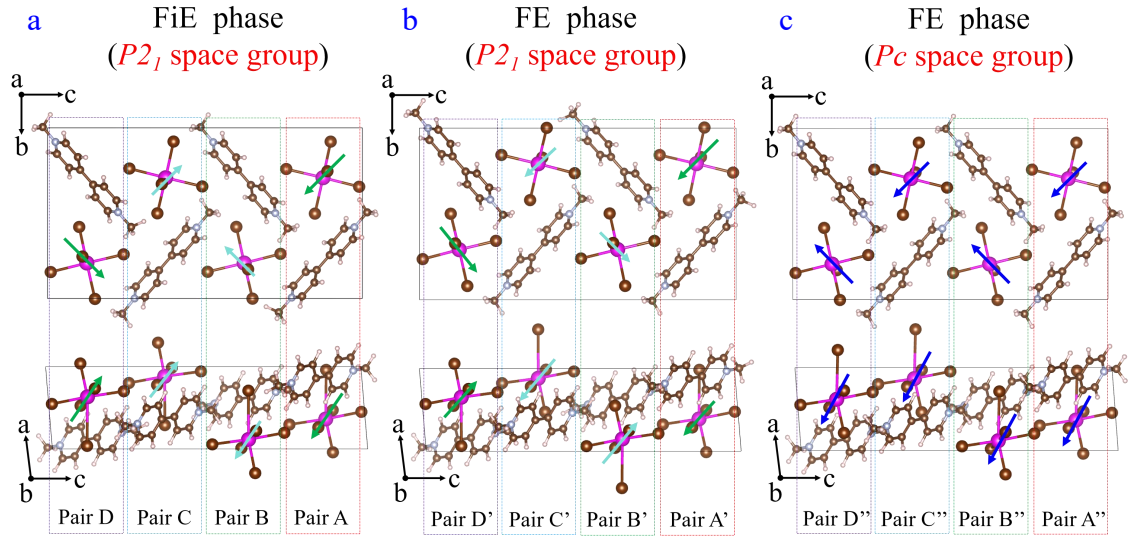

Fig. S10. Schematic illustrations of every pairs (A/A'/A", B/B'/B", C/C'/C" and D/D'/D") for the ferrielectric phase with $P2_1$ space group (a), the ferroelectric phase with $P2_1$ space group (b), and the ferroelectric phase with $Pc$ space group (c). The arrows represent the magnitudes and directions of the dipole moments generated by the movements of Br relative to Sb.

Table S2. The calculated local dipole moments of each (MV)[SbBr$_5$] pairs for the ferrielectric and ferroelectric phases.

| Dipole moment ($|e|$Å) | FiE (II) $P2_1$ space group | III | FE (IV) $P2_1$ space group | ③ | FE (④) $Pc$ space group |
|---|---|---|---|---|---|
| Pair A/A'/A" | (-5.40, 12.72, -23.42) | (-4.95, 12.84, -23.40) | (-5.28, 12.68, -23.46) | (-5.61, 12.74, -23.44) | (-5.38, 12.58, -23.47) |
| Pair B/B'/B" | (-5.28, -12.53, -23.46) | (-5.07, -12.65, -23.42) | (5.29, 12.56, 23.47) | (-5.48, -12.61, -23.47) | (-5.37, -12.58, -23.46) |
| Pair C/C'/C" | (5.28, -12.53, 23.46) | (5.67, -12.55, 23.49) | (-5.29, 12.56, -23.47) | (4.89, -12.61, 23.46) | (-5.38, 12.58, -23.47) |
| Pair D/D'/D" | (5.40, 12.72, 23.42) | (5.57, 12.82, 23.47) | (5.28, 12.68, 23.46) | (5.03, 12.75, 23.44) | (-5.37, -12.58, -23.46) |
| Total Dipole (A+B+C+D) ($|e|$Å) | (0, 0.38, 0) | (1.22, 0.46, 0.14) | (0, 50.48, 0) | (-1.17, 0.27, -0.01) | (-21.50, 0, -93.86) |
| Total Dipole ($\mu C/cm^2$) | (0, 0.33, 0) | (1.06, 0.40, 0.12) | (0, 43.73, 0) | (-1.01, 0.23, -0.01) | (-18.63, 0, -81.35) |
| Berry phase-P ($\mu C/cm^2$) | (0, 0.29, 0) | (1.16, 0.22, 0.37) | (0, 43.96, 0) | (-1.16, 0.17, -0.05) | (-18.48, 0, -81.64) |
| [1 1 1]-P | 0.12 | 1.22 | 17.81 | -0.97 | -39.92 |

| (μC/cm²) | | | | | |

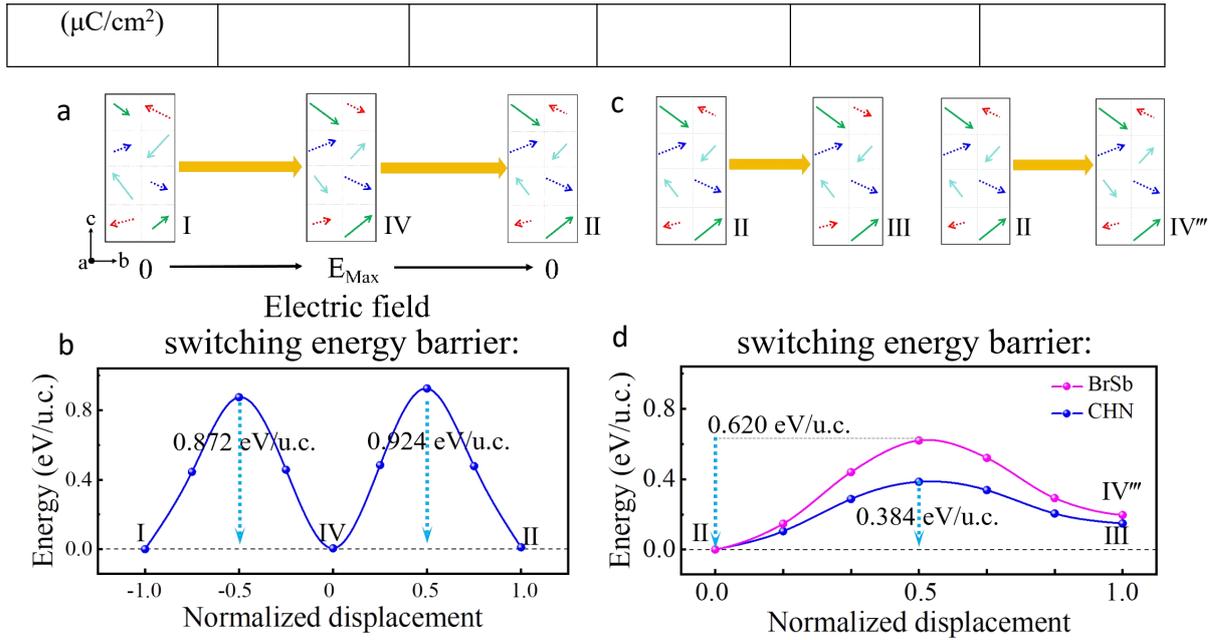

Fig. S11. Possible polarization reversal pathways and the corresponding energy barriers. a, Schematic illustration of a possible (different from the one in the main text) polarization reversal pathway of (MV)[SbBr$_5$] under an electric field: FiE$^{(-)}$ → FE → FiE$^{(+)}$ (I → IV → II), and (b) the corresponding energy barriers. c, Schematic diagrams illustrating the dipole reversal contributed by the inorganic frameworks and the movement of MV cations during the FiE$^{(+)}$ → FE (II → IV); process. d, The dipole reversal energy barriers of the inorganic framework (purple line: II → IV‴) and the MV cations (blue line: II → III) during the FiE$^{(-)}$ → FE process.

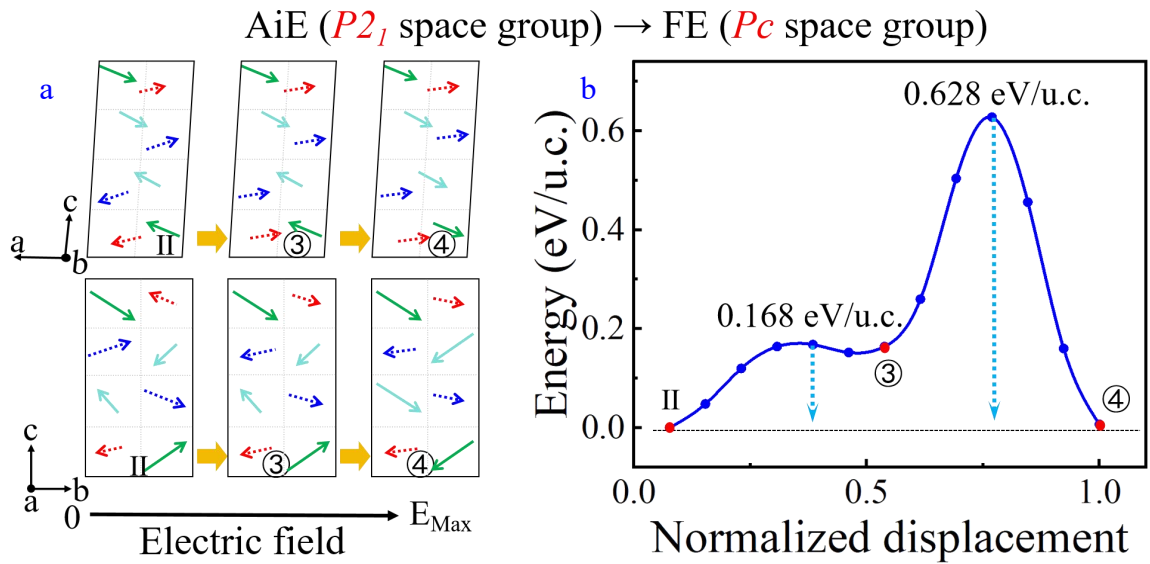

Fig. S12. Another polarization reversal pathway and the corresponding energy barriers. a, The polarization reversal path under an electric field applied along the a-axis is driven by the rotation of dipoles, which induces polarization along both the a-axis and the c-axis, and (b) the corresponding energy barriers.

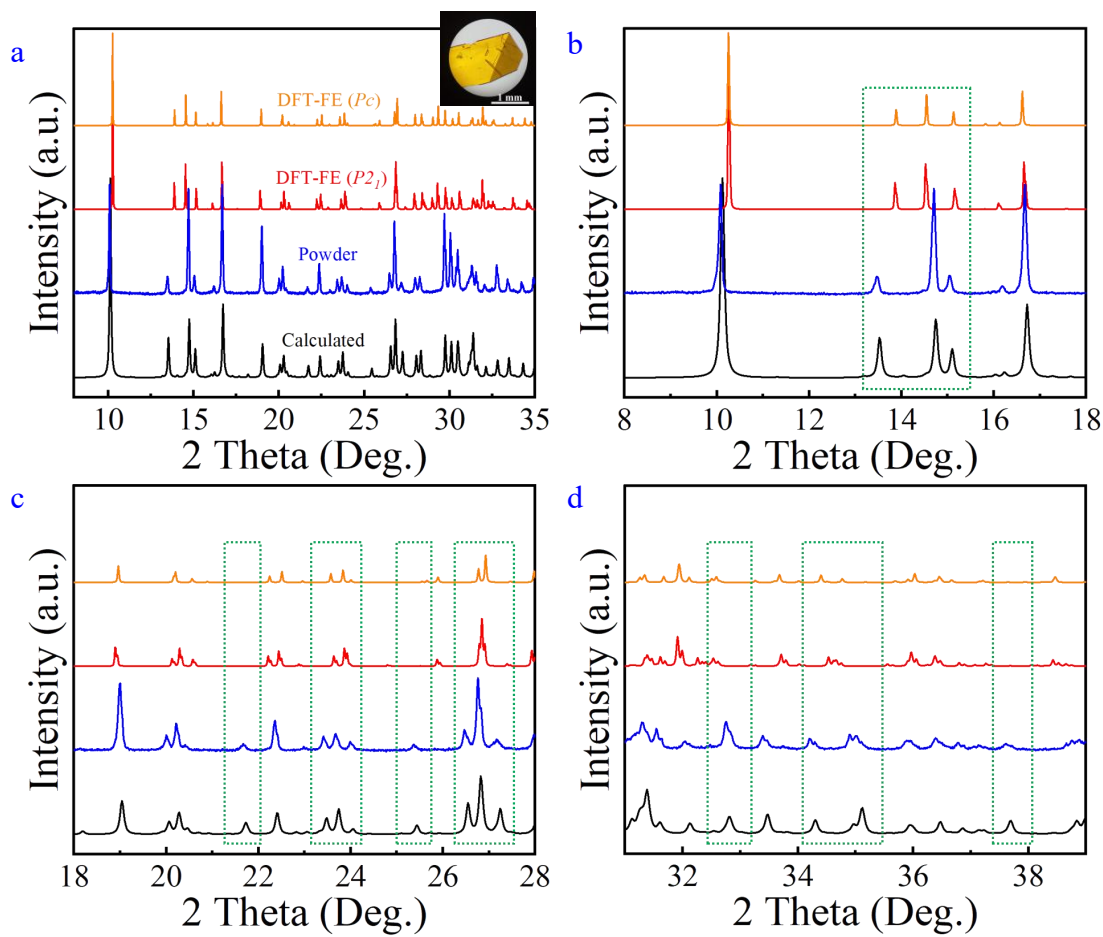

Fig. S13. Further analysis of the (MV)[SbBr$_5$] crystal phase purity. a, Comparison of the calculated XRD pattern for the experimental structure of (MV)[SbBr$_5$], the experimental powder XRD, and the calculated XRD patterns for the two ferroelectric phases from DFT. Inset: Image of the crystal surface observed under a polarized light microscope, showing two distinct regions with single contrast, indicating that both regions are of a single composition. b, c, d, The enlarged detail of the XRD patterns, indicating the high purity of the (MV)[SbBr$_5$] crystal.

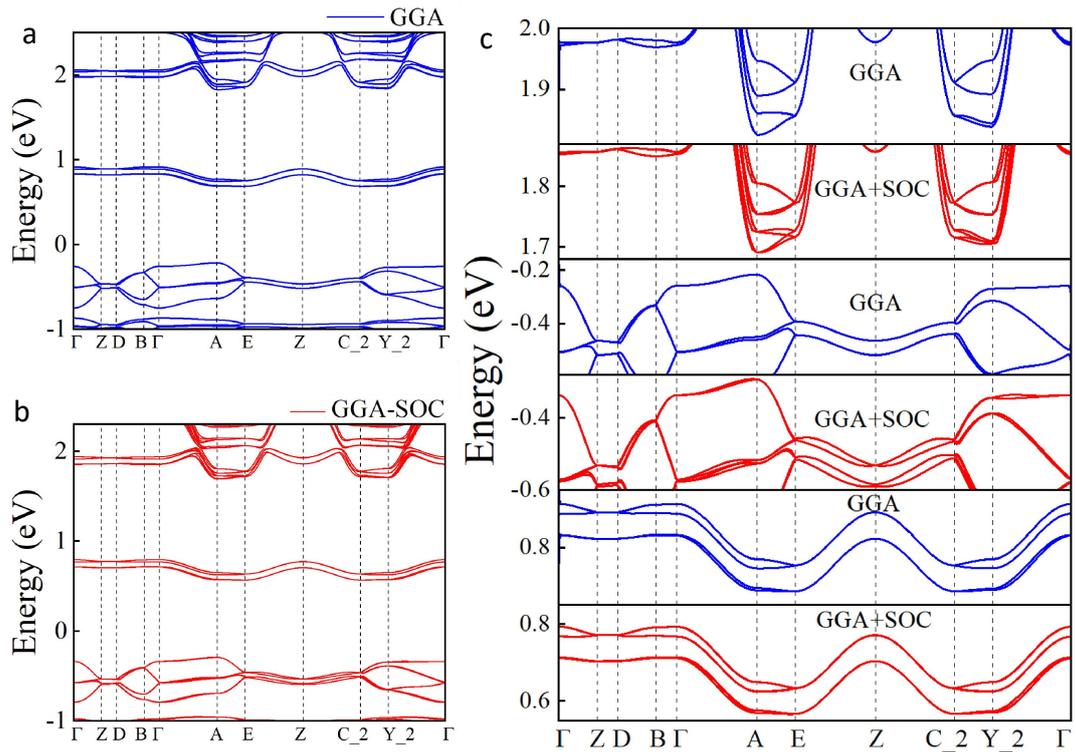

Fig. S14. DFT calculated electronic band structures of (MV)[SbBr$_5$]. a, DFT-GGA + SOC band structure; b, DFT-GGA band structure; c, Local zoom-in of the DFT-GGA + SOC band structure and the DFT-GGA band structure.

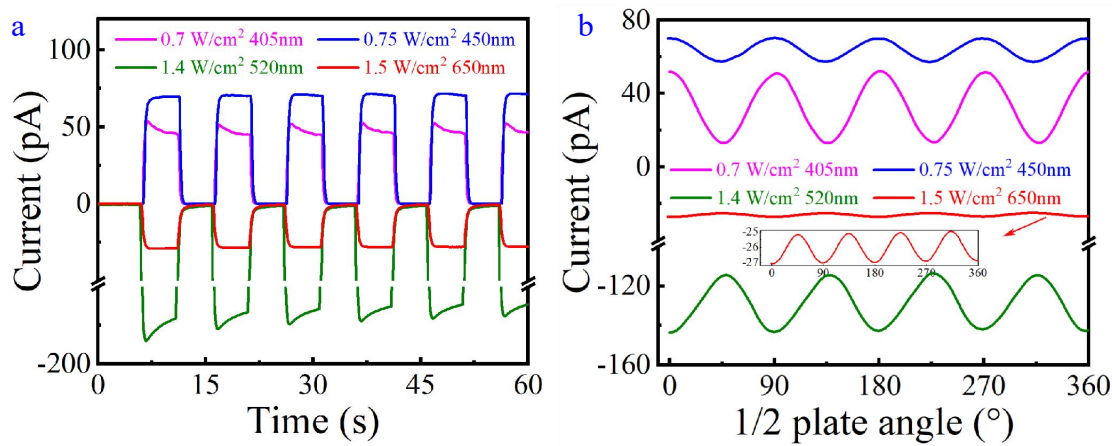

Fig. S15. Photovoltaic responses of (MV)[SbBr$_5$] at different wavelengths. a, I-t responses upon illumination with lights of different wavelengths. b, Room temperature photovoltaic currents under light of different wavelengths vs. λ/2 plate rotation angle.

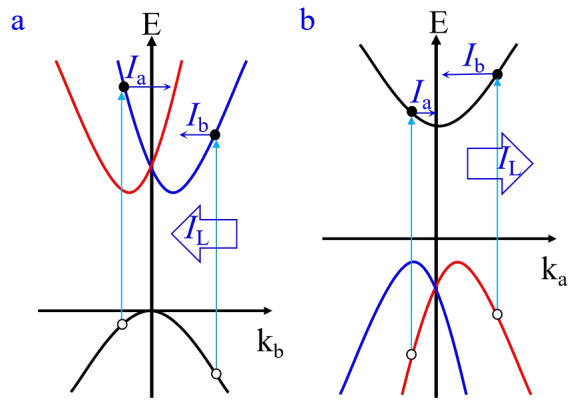

Fig. S16. The influence of valence band and conduction band spin splitting on CPGE photocurrents. a, Under LCP illumination, the spin splitting along the b-axis generates a CPGE photocurrent in the opposite direction along the b-axis; b, Under LCP illumination, the spin splitting along the a-axis generates a CPGE photocurrent in the direction of the a-axis.

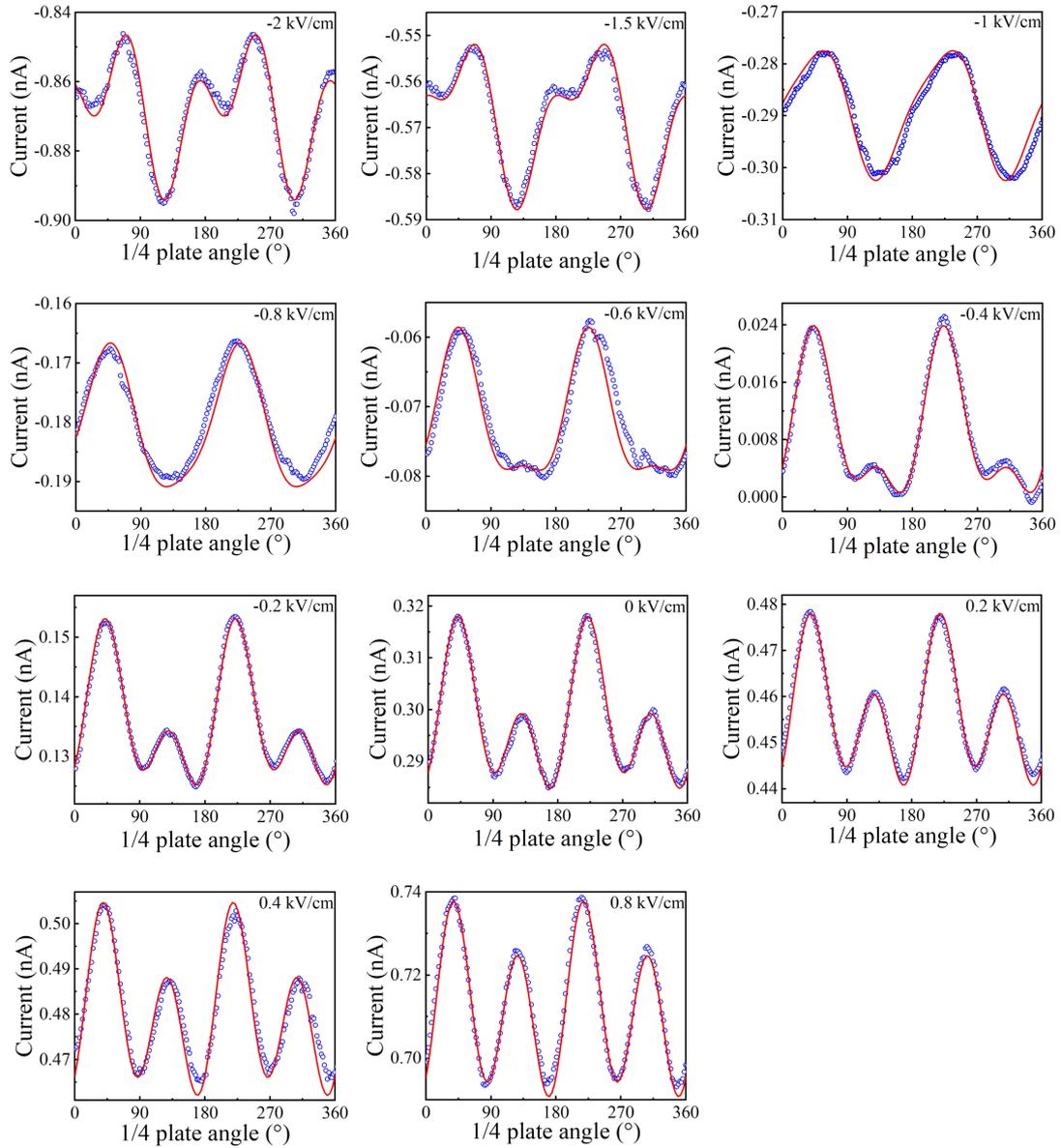

Fig. S17. Circular photogalvanic effect at different bias and the corresponding fitting curve. The blue circulars show the dependence of photocurrent on light polarization under different bias upon 405 nm light illumination. The red line represents the curve fitted according to Eq. 1.

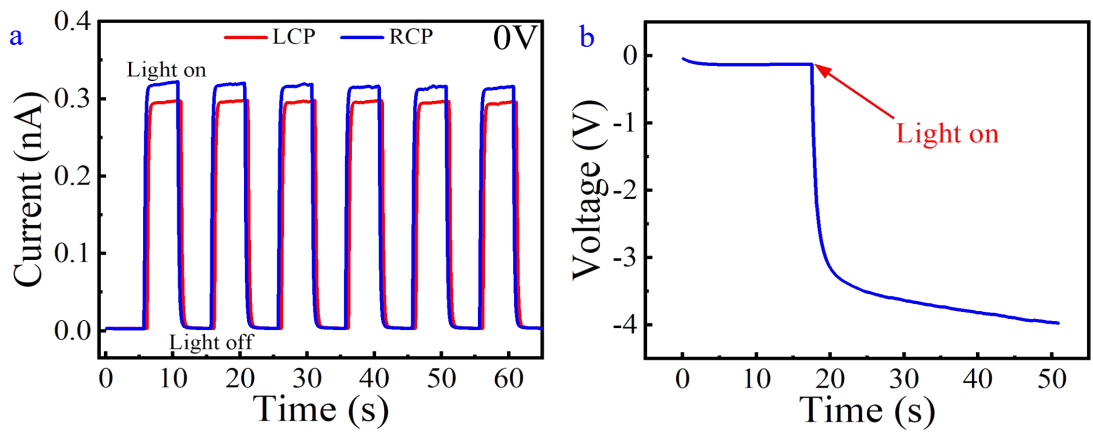

Fig. S18. Photovoltaic responses of (MV)[SbBr$_5$] under circularly polarized lights. a, Photocurrents upon left- (LCP) and right-circularly polarized (RCP) light irradiation at 405 nm with an energy density of 0.7 W/cm$^2$; b, Open-circuit voltage ($V_{oc}$) versus time when the light is turned on.

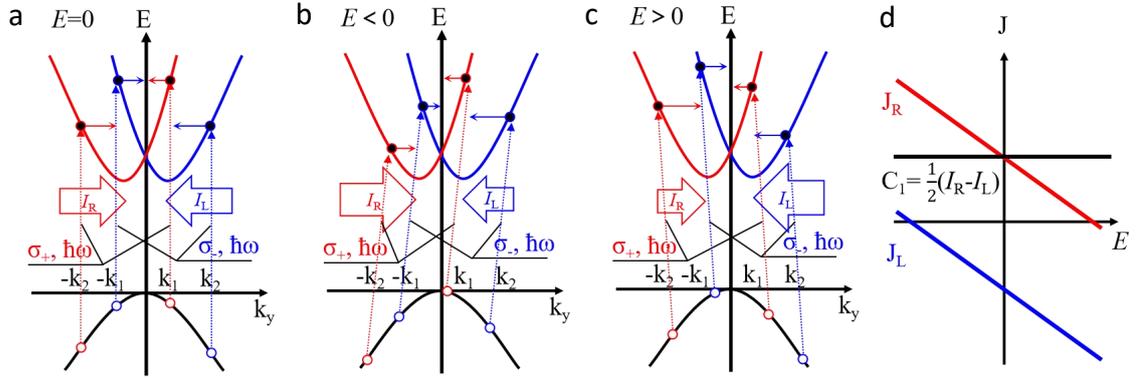

Fig. S19. The effect of bias on the distribution of excited electrons and the CPGE photocurrent. a, The microscopic origin of CPGE arises from the spin splitting of electrons/holes. The RCP/LCP light excitation is represented by dashed lines with red/blue arrows, while the relaxation of the excited electrons is indicated by solid lines with red and blue arrows. Under excitation by RCP (LCP) light at a frequency of ω, energy and momentum conservation only allow transitions at two $k_y$ values labeled $-k_2$ and $k_1$ ($-k_1$ and $k_2$). The asymmetric distribution of electrons in k-space results in a non-zero group velocity of the electrons, generating photocurrent $I_R$ ($I_L$). b and c, The effect of external bias on CPGE. The external bias provides additional momentum to the photoexcited electrons, thereby altering the photocurrent generated under right/left circularly polarized light excitation. However, since the CPGE coefficient $C_0 = (I_R - I_L)/2$, the value of $C_1$ should remain constant. d, The correlation between the CPGE currents $I_R$ and $I_L$ and the CPGE coefficient $C_1$ under external bias.

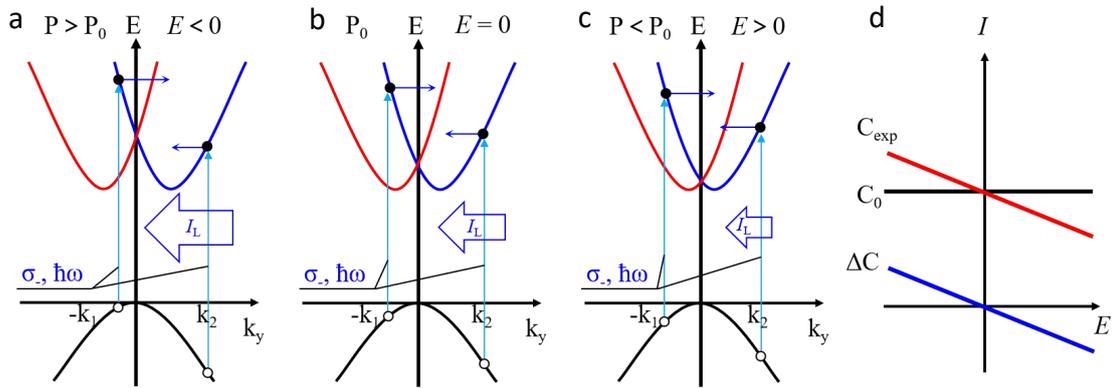

Fig. S20. The effect of bias on polarization, band spin splitting, and CPGE photocurrent. a, Applying a bias along the polarization direction increases the polarization value, enhances the spin splitting of the energy bands, and increases $I_L$. b, At zero bias, the polarization value is denoted as $P_0$. c, Applying a bias opposite to the polarization direction decreases the polarization value, reduces the spin splitting of the energy bands, and decreases $I_L$. d, The experimentally expected CPGE coefficient $C_{exp}$ due to the effect of external bias on polarization, and its relationship with the initial CPGE coefficient $C_0$, correlates with the compensation photocurrent $\Delta C$.

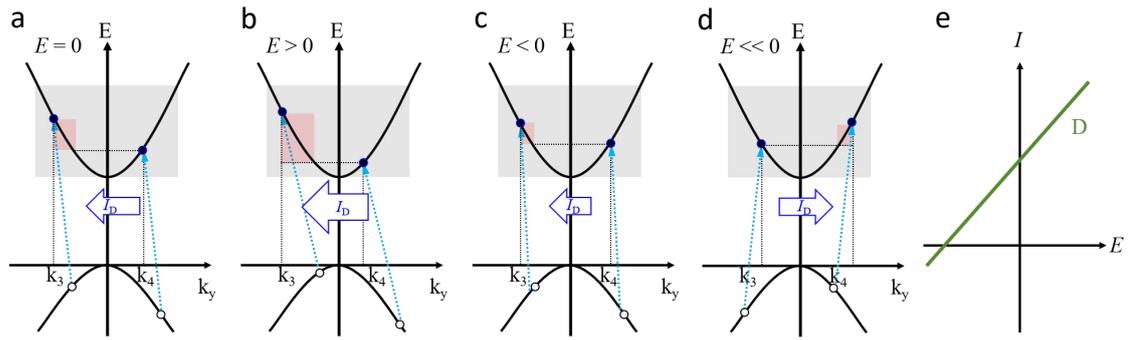

Fig. S21. The effect of bias on the distribution of excited electrons and the LPDE photocurrent. At (a) $E=0$, (b) $E>0$, (c) $E<0$, and (d) $E<<0$, explore the influence of bias on LPDE photocurrent. e, The variation in LPDE photocurrent with bias.

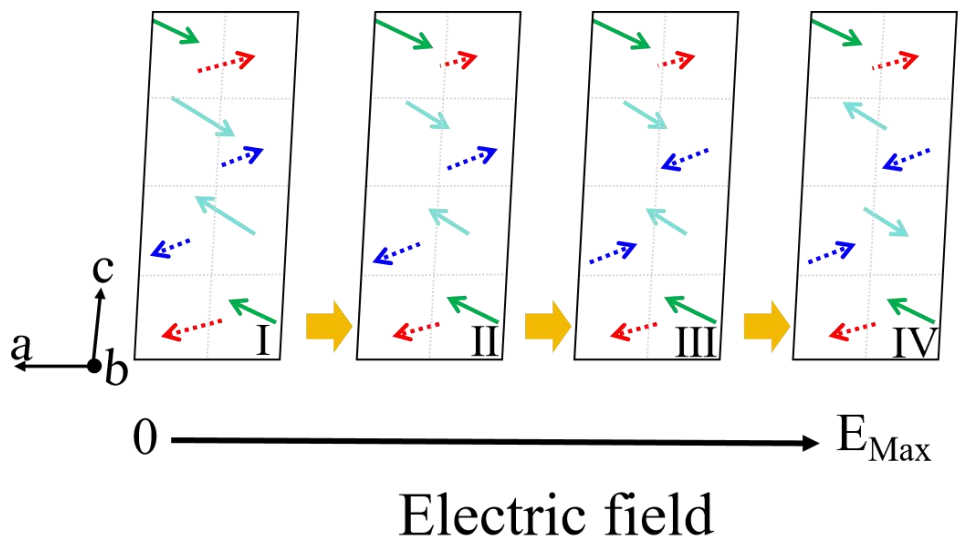

Fig. S22. The dipole evolution in the *ac*-plane under an electric field applied along the *b*-axis shows that the local dipole moment components along the *a*-axis and *c*-axis are consistently canceled by oppositely oriented dipole moments.

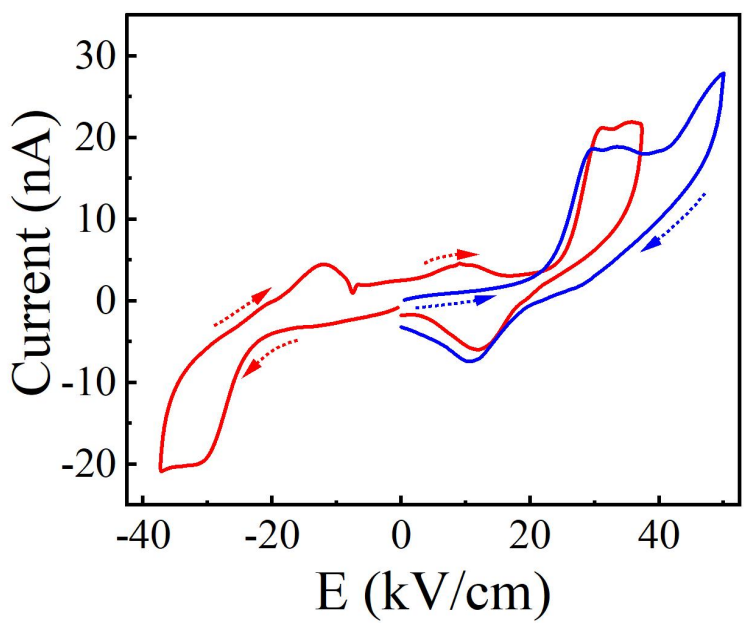

Fig. S23. The current-electric field curve obtained using the Keithley 2636B source meter.

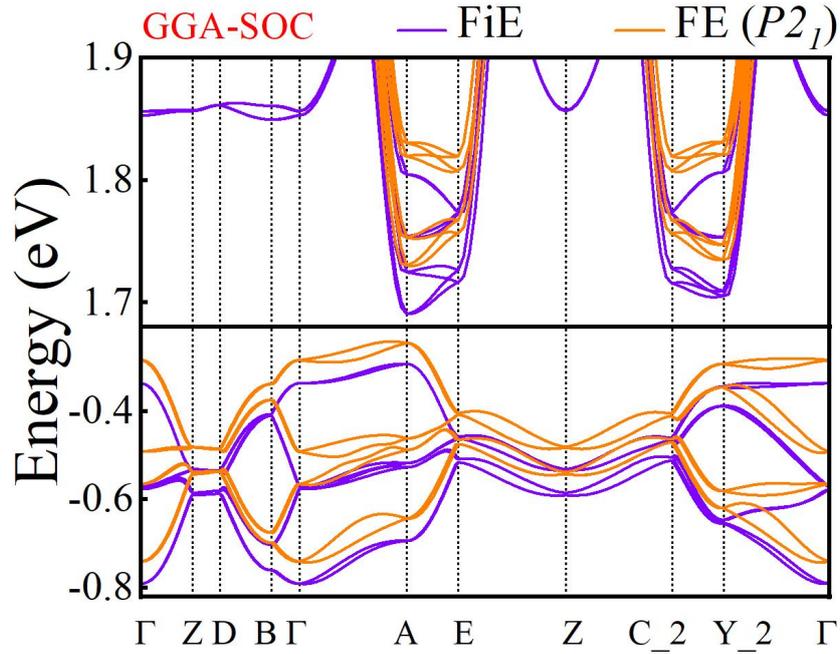

Fig. S24. Comparison of the bands between the ground state FiE and the electric-field-induced ferroelectric state ($P2_1$ phase) considering SOC.

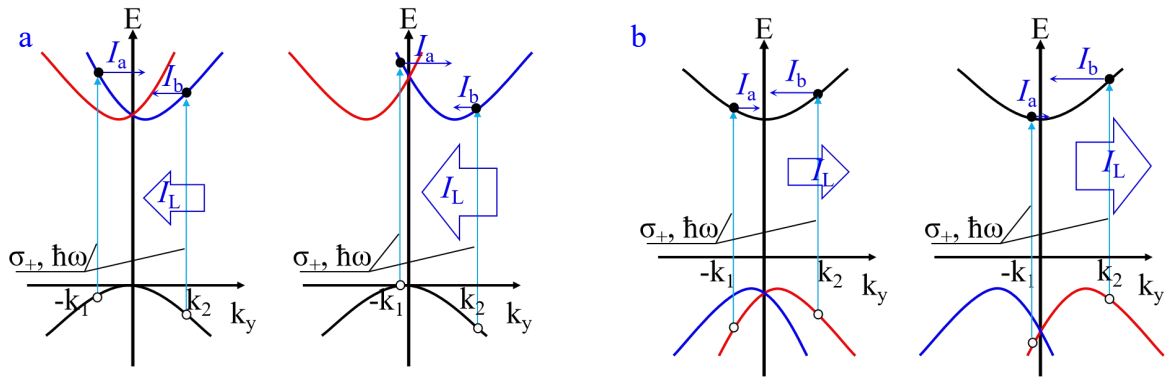

Fig. S25. a and b, show a comparison of the CPGE photocurrent in the ground state FiE and the electric-field-induced ferroelectric state FE, driven by spin splitting dominated by the conduction band and the valence band, respectively, under LCP light excitation.